\documentclass[aps,prb,twocolumn,showpacs,10pt]{revtex4-1}
\usepackage{graphicx}
\usepackage{amssymb}
\usepackage{amsmath,color,mathrsfs}
\usepackage{bbold}
\usepackage{comment}
\usepackage[dvipsnames]{xcolor}
\usepackage{bm}
\pdfoutput=1
\usepackage{hyperref}
\hypersetup{urlcolor=blue, colorlinks=true, citecolor=blue, linkcolor=blue}

\newcommand{\up}{\uparrow}
\newcommand{\down}{\downarrow}

\begin{document}
	
	\title{Conductance of fractional Luttinger liquids at finite temperatures}
	\author{Pavel P. Aseev}
	\author{Daniel Loss}
	\author{Jelena Klinovaja}
	\affiliation{Department of Physics, University of Basel, Klingelbergstrasse 82, CH-4056 Basel, Switzerland}
	
	\date{\today}
	
	\begin{abstract}		
		We study the electrical conductance in single-mode quantum wires with Rashba spin-orbit interaction subjected to externally applied magnetic fields in the regime in which the ratio of spin-orbit momentum to the Fermi momentum is close to an odd integer, so that a combined effect of multi-electron interaction and applied magnetic field leads to a partial gap in the spectrum. We study how this partial gap manifests itself in the temperature dependence of the fractional conductance of the quantum wire. We use two complementing techniques based on bosonization: refermionization of the model at a particular value of the interaction parameter and a semiclassical approach within a dilute soliton gas approximation of the functional integral. We show how the low-temperature fractional conductance can be affected by the finite length of the wire, by the properties of the contacts, and by a shift of the chemical potential, which takes the system away from the resonance condition. We also predict an internal resistivity caused by a dissipative coupling between gapped and gapless modes.	
	\end{abstract}

	\maketitle
	
	\section{Introduction}

Electron systems, in which excitations with non-Abelian statistics or fractional charge, such as Majorana fermions or parafermions, may exist have attracted much attention in recent years, since they are of great interest both for understanding of the underlying physics and for applications in topological quantum computing\cite{KitaevUFN2001,KitaevAnnPhys2006, NayakRevModPhys2008, AliceaRepProgPhys2012, FendleyStatMech2012, AliceaFendleyAnnuRev2016}. 
In particular, topological phases were also predicted in one-dimensional (1D) helical liquids manifesting spin-filtered transport \cite{SatoPRL2010, LeeBruderPRB2007, KlinovajaPRL2011, KlinovajaPRB2012, MengKlinovajaPRB2014, KlinovajaLossPRB2015, SchliemannRevModPhys2017, QuayNatPhys2010, BernardesPRL2007, GovernalePRB2002, CalsaveriniPRB2008, PerroniJPhys2007, AbaninPRL2006, GovernaleBoesePRB2002},  particularly in a single-mode quantum wire with Rashba spin-orbit interaction and Zeeman magnetic field \cite{StredaPRL2003, BrauneckerPRB2010, OregPRB2014}. If the ratio of the spin-orbit momentum $k_{so}$ to the Fermi momentum of the wire $k_F$ is an odd integer, $\gamma_c = 2n+1$, multi-electron processes involving large momentum transfer may lead to an opening of an energy gap, and a fractional helical liquid state can appear \cite{OregPRB2014}. This state manifests itself as a fractional two-terminal conductance at zero temperature $G(T=0) = 2G_0/(1+\gamma_c^2)$, where $G_0 = e^2/h$, and also reveals itself in optical conductivity\cite{MengPRB2014}, tunneling density of states\cite{MengPRB2014} and shot noise\cite{CornfeldPRB2015}, which allows one to observe this state in principle using state-of-the-art experimental techniques~\cite{SchellerPRL2014, HeedtNatPhys2017, KammhuberNatComm2017}. 
	The fractional helical electron systems are also considered to be one of the ingredients for possible experimental realization of fractional bound states  
	and parafermions \cite{BarkeshliPRB2012,OregPRB2014,LindnerPRX2012,KlinovajaLossPRL2014}.

		\begin{figure}		
		\includegraphics[width=\linewidth]{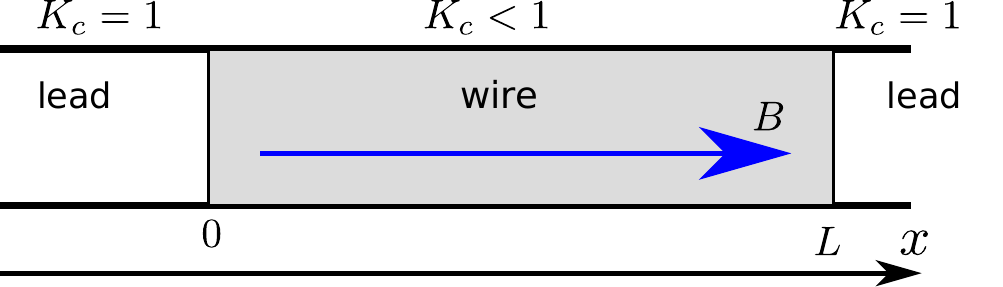}
		\caption{Single-band quantum wire with leads attached from the left and right side at positions $x=0$ and $x=L$, respectively. A magnetic field $B$ is applied along the wire axis in $x$ direction. Repulsive electron-electron interactions in the quantum wire are described within a Luttinger liquid model with  interaction parameter $K_c<1$. The leads are assumed to be non-interacting with $K_c=1$. }
		\label{fig:geometry}
	\end{figure}	
	
In this paper we study the temperature dependence of the conductance of a single-band quantum wire ~(see Fig.~\ref{fig:geometry}) in the fractional regime~\cite{OregPRB2014} in which the electrons form a fractional helical Luttinger liquid with spin and charge sectors being locked by a sine-Gordon potential. Using two complementing techniques, refermionization at a special value of interaction parameter and a semiclassical expansion around the static soliton solutions~\cite{Rajaraman1982,GoldstoneJackiwPRD1975,Aseev2017}, we describe how the  electric conductance depends on the length of the quantum wire, on temperature, and  on the shift of the chemical potential away from the resonance values such that the ratio $k_{so}/k_F$ is slightly tuned away from integer values assumed above. Moreover, since transport properties in low-dimensional systems are strongly affected by the attached leads, we also study how the conductance depends on the properties of the contacts. 
	
 We also predict a resistivity mechanism caused by electron-electron interactions. At non-zero temperatures soliton excitations   carrying electric charge can be activated. These solitons are coupled to gapless excitations which leads to an Ohmic-like friction for these solitions and thus to a temperature-dependent resistivity. 
 While this mechanism resembles the one recently described for Rashba nanowires with weakly interacting electrons and a partial gap induced by magnetic Zeeman field\cite{SchmidtPRB2014}, previous studies cannot be used to take into account effects of strong electron-electron interactions, which is necessary for the formation of a fractional Luttinger liquid considered here.

	The outline of the paper is as follows. In~Sec.~\ref{sec:model} we introduce a model of a fractional Luttinger liquid. We discuss the results obtained by refermionization of the bosonized model at a particular value of the interaction parameter in Sec.~\ref{sec: refermionization}. In Sec.~\ref{sec:semiclassical} we study the electric conductance using semiclassical expansions around static soliton configurations. Finally, we conclude  with ~Sec.~\ref{sec:results}, where we summarize our results.

	\section{The model\label{sec:model}}
	
	\begin{figure*}[t]	
		\includegraphics[width=\linewidth]{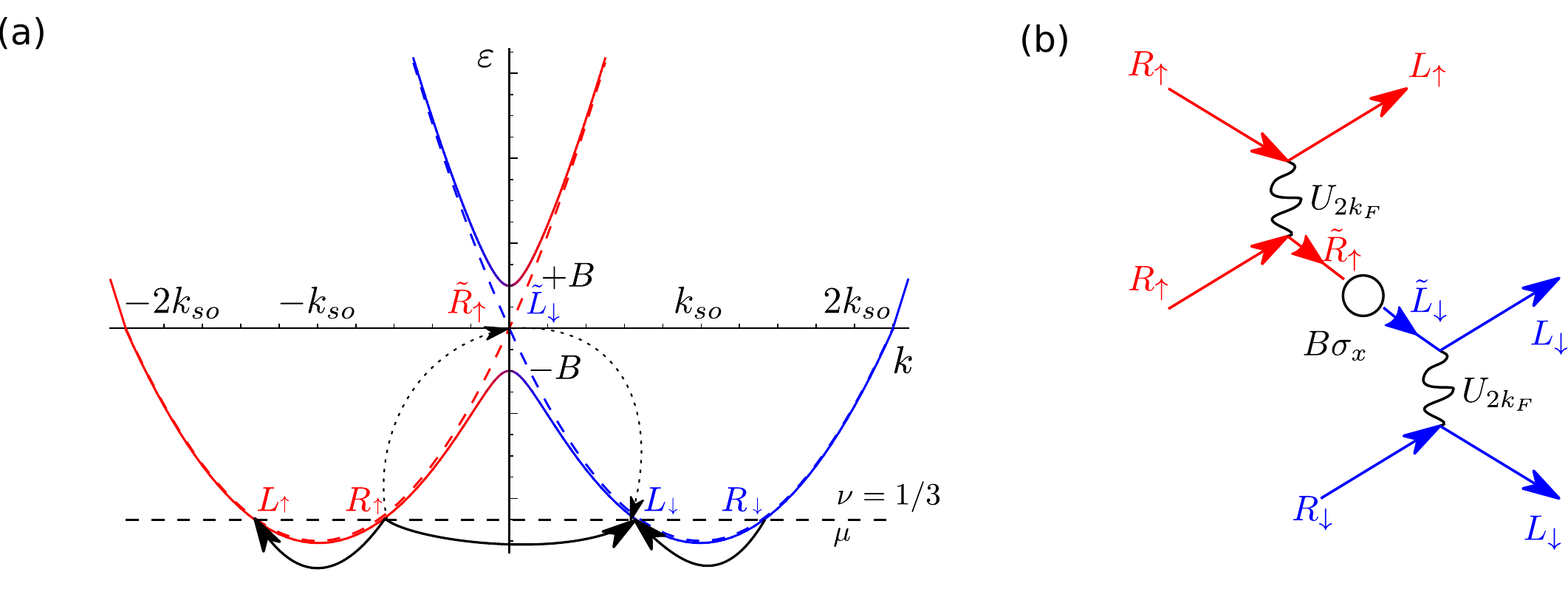}
		\caption{(a)  Spectrum of a quantum wire consists of two spin-orbit split bands. A helical gap is opened by the Zeeman term at $k=0$ if a magnetic field is applied perpendicular to the SOI vector, which determines the quantization axis. A horizontal dashed line shows the position of the Fermi energy at filling factor $\nu=1/3$, at which the three-particle scattering (shown by black arrows) conserves momentum. Dotted arrows depict virtual transitions to/from states with a momentum close to zero. 		
			(b) Diagram for the multi-particle scattering process ($n=1$) corresponding to the interaction term given by Eq.~(\ref{eqn:interaction-term}). The right (left)-moving electrons with momentum close to the Fermi points $\pm 2k_{so}/3$, $\pm 4k_{so}/3$ are labeled as $R_s$ ($L_s$) with the spin label $s=\uparrow,\downarrow$, whereas right (left)-moving electrons with momentum close to $k=0$ are labeled by $\tilde{R}$, $\tilde{L}$. The interplay between electron-electron interactions at characteristic momenta $2k_F \equiv 2k_{so}/3$ (wavy lines) and magnetic field $B$ (circle vertex) results in a partial gap at $\nu=1/3$.  
	}
		\label{fig:spectrum}
	\end{figure*}	
	
	We consider a spinful single-band quantum wire of length $L$ aligned in  $x$-direction~(see Fig.~\ref{fig:geometry}) The Rashba spin-orbit interaction (SOI) of  strength $\alpha$ sets the spin quantization axis to be along the $y$-axis. A magnetic field $B$ is applied along the wire (below, for the sake of conciseness, we absorb the electron $g$-factor and Bohr magneton $\mu_B$ into the symbol $B$). The single-particle Hamiltonian is given by $H=H_0 + H_B$, 
	\begin{align}
	&H_0 = -\frac{\hbar^2 \partial_x^2}{2m} - \mu -i  \alpha \sigma_y \partial_x,
	\label{eqn:Hamiltonian0}\\
	&H_B = B\sigma_x,
	\label{eqn:Hamiltonian}
	\end{align}
where  $\sigma_\nu$ (with $\nu=x,y,z$) are the Pauli matrices acting on spin and $m$ is an effective mass of electrons in the quantum wire. The SOI leads to a shift of the momentum $k$ of the free-electron parabolic spectrum (see Fig.~\ref{fig:spectrum}a). The spin-up (along the $y$-axis) electronic dispersion is shifted to the left by the SOI momentum $k_{so} = m\alpha$, while the dispersion of electrons with the opposite spin is shifted to the right by $k_{so}$. The chemical potential $\mu$ is  measured from the crossing point between spin-up and spin-down bands at $k=0$~(see Fig.~\ref{fig:spectrum}). The uniform  magnetic field (Zeeman term) applied along the wire opens a (helical) gap in the spectrum near $k=0$. In what follows, we assume that the Zeeman energy $B$ is small in comparison to  the SOI energy $\varepsilon_{so} = m\alpha^2/2$, $B\ll \varepsilon_{so}$. Here and below, we take $\hbar = 1$, $e=1$, and $k_B=1$, restoring physical units when necessary.  
	
In this paper we consider effects of electron-electron interactions, which can be taken into account by means of the Luttinger liquid formalism~\cite{GiamarchiBook}.  We note that multi-electron processes involving large-momentum transfer (backscattering terms) may cause an opening of an energy gap at proper values of $\mu$~\cite{OregPRB2014,SchmidtPRB2016}.
If $\mu$ is below the gap at $k=0$, there are four Fermi points with wave vectors $\pm(k_{so} \pm k_F)$, and we define four fermionic fields: right-moving modes ($R_\uparrow$, $R_\downarrow$) and left-moving modes ($L_\uparrow$, $L_\downarrow$), such that the electron annihilation operators  can be represented as 	
	\begin{align*}
	&\psi_\uparrow(x) = R_\uparrow(x)e^{i(-k_{so}+k_F)x} + L_\uparrow(x)e^{-i(k_{so}+k_F)x},\\
	&\psi_\downarrow(x) = R_\downarrow(x)e^{i(k_{so}+k_F)x} + L_\downarrow(x)e^{i(k_{so}-k_F)x}.
	\end{align*}
Following~Refs.~\onlinecite{OregPRB2014, KanePRL2002}, we focus on the back-scattering interaction term~(see Fig.~\ref{fig:spectrum}):
		\begin{align}
	\mathcal{O}^B_n = g_B^{(n)}\left(L^\dag_{\uparrow} R_{\uparrow} \right)^n R_{\uparrow}L_{\downarrow}^\dag \left(L_{\downarrow}^\dag R_{\downarrow} \right)^n + H.c.
	\label{eqn:interaction-term}
	\end{align}
The interaction conserves the spin and it also conserves momentum at the filling factor $\nu \equiv k_F/k_{so} = 1/(2n+1)$, which corresponds to the following positions of the chemical potential, $\mu_n = -\varepsilon_{so} + \varepsilon_{so}/(2n+1)^2$. We note that such an interaction term [see Eq.~(\ref{eqn:interaction-term})] is a combined result of a magnetic field and electron-electron interaction. 
Throughout the paper, we mainly focus on the case $n=1$. In this case,	$\mathcal{O}^B_1$ is generated at the second order in the bare interaction strength [see Fig.~(\ref{fig:spectrum}b)], so that $g_B\equiv g_B^{(1)}$ has the following structure,
	\begin{align}
	g_B\sim BU_{2k_F}^2 \left\langle \tilde{R}_{\up} \tilde{R}_{\up}^\dag  \right\rangle_0 \left\langle \tilde{L}_{\down} \tilde{L}_{\down}^\dag  \right\rangle_0,
	\label{eqn:gB}
	\end{align} 	
	where $U_q$ is the electron-electron interaction potential and  $\tilde{R}_s, \tilde{L}_s$ denote right- and left-moving electrons with momentum close to zero, and $\langle\dots\rangle_0$ means thermodynamic average for the Hamiltonian $H_0$ given by Eq.~(\ref{eqn:Hamiltonian0}).

At temperatures and magnetic fields much lower than the Fermi energy, $T, B\ll \varepsilon_F \simeq m\alpha^2/\left[2(2n+1)^2\right]$, one can linearize the spectrum at Fermi points and follow the standard bosonization procedure~\cite{GiamarchiBook}. The bosonized Euclidean action is given by~\cite{OregPRB2014,MengPRB2014}
	\begin{align}
	&S_E= \frac{1}{2\pi}\int dxd\tau\ \phi_c \left[-\partial_x \frac{v_c(x)}{K_c(x)}\partial_x - \frac{1}{v_c(x) K_c(x)}\partial_\tau^2 \right]\phi_c \nonumber\\
	&+\frac{1}{2\pi}\int dxd\tau\ \theta_\sigma \left[-\partial_x v_\sigma(x)K_\sigma(x)\partial_x - \frac{K_\sigma(x)}{v_\sigma(x)}\partial_\tau^2 \right]\theta_\sigma \nonumber \\
	&+\int dxd\tau\; \frac{\Delta_0}{2\pi a} \cos\left(\sqrt{2} \left[ \gamma_c \phi_c(x,\tau) + \theta_\sigma(x,\tau) \right] \right),
	\label{eqn:Matsubura-rho-sigma-action}	
	\end{align}
	where the bosonic fields $\phi_c$ and $\theta_\sigma$ relate to the integrated charge  and spin density current, respectively, and $\tau$ is the Matsubara time. The effective velocities of the charge and spin excitations, $v_c$ and $v_\sigma$, $v_{c,\sigma}(x) = v_F/K_{c,\sigma}(x)$ are related to the Fermi velocity $v_F= k_F/m$
and corresponding Luttinger liquid (LL) parameters $K_{c,\sigma}$. The short-distance cutoff parameter is determined by $a\sim k_{so}^{-1}$. The amplitude of the sine-Gordon term describing the locking of charge degrees of freedom is given by $\Delta_0 = g_B^{(n)}/(2\pi a)^{2n}$. At $n=1$, it can be estimated as $\Delta_0 = BU_{2k_F}^2/v_F^2 $.  The interaction term, given by Eq.~(\ref{eqn:interaction-term}), results in $\gamma_c = 2n+1$.  We note that the model with $\gamma_c=1$ and $\Delta_0 = B$ corresponds to Rashba quantum wires with the Fermi level located at the middle of the helical gap. 
	
To take metallic leads into account, we assume similarly to~Ref.~\onlinecite{MaslovStonePRB1995} that the LL parameters $K_{\nu}$ depend on the coordinate $x$, so that the interaction vanishes in the leads, $K_{\nu} = 1$~(see~Fig.~\ref{fig:geometry}). Below in numerical estimations we assume that the Fermi level is fine-tuned to a specific value, so that $k_{so}=3k _F$, {\it i.e.}, $n=1$, $\gamma_c = 3$. We also focus on the case $K_\sigma=1$, assuming spin-rotational symmetry of the electron-electron interaction. In general, we assume that $\Delta_0(x) \propto BU_{2k_F}^2$ depends on the coordinate, since both the interaction potential and the magnetic field are, in general, non-uniform. In this paper we consider two limiting cases: (i) the gap vanishes in the leads abruptly $\Delta_0(x) = \Delta_0\Theta(x)\Theta(L-x)$, where $\Theta(x)$ is the Heaviside step-function; (ii) the gap $\Delta_0(x)$ varies adiabatically in space, and its spatial dependence is modeled as
\begin{align}
\Delta(x) = \frac{\Delta}{2}\left( \tanh \frac{x}{l} + \tanh \frac{L-x}{l} \right),
\label{eqn:gap-profile}
\end{align}
where $l$ is a characteristic length over which the electron-electron interaction switches on.

The scaling dimension of the sine-Gordon term in Eq.~(\ref{eqn:Matsubura-rho-sigma-action}), is given by~\cite{MengPRB2014, OregPRB2014} $D = (\gamma_c^2 K_c + K_\sigma^{-1})/2$, and this term is relevant in the renormalization group (RG) sense for $D<2$, {\it i.e.}, $K_c < (4-1/K_\sigma)/ \gamma_c^2$. For $\gamma_c=3$ and $K_\sigma=1$, this leads to $K_c < 1/3$. The gap can be estimated as
	\begin{align}
	\Delta = \Delta_0\left( \frac{l_c}{a} \right)^{1-D},
	\label{eqn:1st-RG}
	\end{align}
	where 
	$l_c$ is a correlation length given by
	\begin{align}
	l_c = \min \left\{ L, \frac{\hbar v_F}{T}, \frac{\hbar v_F}{\Delta} \right\}.
	\end{align}
	At low temperatures $T$ and for long quantum wires, $l_c$ is determined by the gap itself, resulting in
	\begin{align}
	\Delta_\infty = \Delta_0 \left( \frac{\hbar v_F}{\Delta_0 a} \right)^{(1-D)/(2-D)}.
	\label{eqn:gap}
	\end{align}	
We note that in the limit of strong electron-electron repulsion, $K_c \to 0$ ($D \to 1/2$), we have $\Delta_\infty \propto B^{2/3}$, which is in agreement with the recent study of Rashba wires using Wigner crystal theory and density matrix renormalization group (DMRG) techniques~\cite{SchmidtPRB2016}.
	
At higher temperatures, $T > \Delta_\infty$, Eq.~(\ref{eqn:1st-RG}) yields the following temperature dependence of the gap:
\begin{align}
 \Delta = \Delta_\infty^{2-D} T^{D-1}.
 \label{eqn:1st-RG-b}
\end{align}
Similarly, in short wires, $L < \hbar v_F/\Delta_\infty$, Eq.~(\ref{eqn:1st-RG}) can be rewritten as
\begin{align}
\Delta = \Delta_\infty^{2-D} \left( \frac{\hbar v_F}{L} \right)^{1-D}.
\label{eqn:1st-RG-c}
\end{align}

For further discussion, it is convenient to introduce new bosonic variables $\phi_1$, $\phi_2$, $\theta_1$, and $\theta_2$ related to the standard bosonic fields in the LL model by the following canonical transformation:
	\begin{align}
	  &\phi_1 =\; \frac{\gamma_c \phi_c+\theta_\sigma}{\lambda},\quad
 	  \phi_2 =\; \frac{-\phi_c+\gamma_c\theta_\sigma}{\lambda},\\
 	  &\theta_1 =\; \frac{\gamma_c \theta_c - \phi_\sigma}{\lambda},\quad
 	  \theta_2 =\; -\frac{\theta_c + \gamma_c\phi_\sigma}{\lambda}\, ,
	\end{align}
	where $\lambda = \sqrt{\gamma_c^2+1}$. In terms of the new variables the Euclidean action $S_E=S_1 + S_2 + S_{12}$ consists of the sine-Gordon action $S_1$ describing gapped modes,
	\begin{multline}
	S_1 = -\frac{1}{2\pi v_F}\int dxd\tau\; \left\{ \phi_1\left[  -\partial_\tau^2 -  \partial_x v_{1}^2 \partial_x \right]\phi_1\right.\\ \left.
	 +\omega_0^2\cos \left(\sqrt{2}\lambda \phi_1\right)\right\},
	 \label{eqn:action-gapped}
 	\end{multline}
	a standard LL action $S_2$ describing gapless modes,
	\begin{align}
S_2 = -\frac{1}{2\pi v_F}\int dxd\tau\;  \phi_2\left[  - \partial_\tau^2  -  \partial_x v_{2}^2 \partial_x  \right]\phi_2,
	\end{align}
	and a coupling $S_{12}$ between gapless and gapped modes,
	\begin{align}
S_{12} = -\frac{1}{2\pi v_F}\int dxd\tau\;  \phi_1\left[ 2\partial_x v_{12}^2 \partial_x \right]\phi_2.	
\label{eqn:action-coupling}
	\end{align}
Here, we use notations $\omega_0^2 = \Delta_0 v_F/a$ and $v_{i} = v_F/K_{i}$, where the parameters $K_{1}$, $K_{2}$, $K_{12}$ are related to the LL parameter $K_c$ (with $K_\sigma=1$) as
	\begin{align}
	&\frac{1}{K_{1}^{2}} = \frac{\gamma_c^2 K_c^{-2} +1}{\lambda^2}, \, \frac{1}{K_{2}^{2}} = \frac{\gamma_c^2 + K_c^{-2}}{\lambda^2},\\
	& \frac{1}{K_{12}^{2}} = \gamma_c\frac{K_c^{-2}-1}{\lambda^2}.
	\label{eqn:vij}
	\end{align}
As a result, the charge current is given by
	\begin{align}
	j_c = \frac{\sqrt{2}e}{\pi} \dot{\phi}_c = \frac{\sqrt{2} e}{\pi \lambda}\left[ \gamma_c \dot{\phi}_1 - \dot{\phi}_2 \right].
	\end{align}
The conductance at zero voltage bias can be extracted from the Matsubara Green functions via the Kubo formula\cite{MaslovStonePRB1995},
\begin{multline}
	G = \frac{2e^2}{\pi^2 \lambda^2}\lim\limits_{\bar{\omega}\to 0}\bar{\omega}\left(\gamma_c^2 \left\langle \phi_1(x,\bar{\omega})\phi_1(x,-\bar{\omega})\right\rangle \right. \\ \left. +  \left\langle \phi_2(x,\bar{\omega})\phi_2(x,-\bar{\omega})\right\rangle +\gamma_c\left\langle \phi_1(x,\bar{\omega})\phi_2(x,-\bar{\omega}) \right\rangle \right. \\ \left.+\gamma_c\left\langle \phi_2(x,\bar{\omega})\phi_1(x,-\bar{\omega}) \right\rangle\right),
	\label{eqn:Kubo}
	\end{multline}
	where $\langle \dots \rangle$ means thermodynamic average, $\bar{\omega}$ is the Matsubara frequency, and the Fourier transform is defined as $\phi_i(x,\bar{\omega}) = \int_0^{1/T} d\tau\; \phi_i(x,\tau) e^{i\tau\bar{\omega}} $.

		\begin{table}
		\begin{tabular}{|c||c|c|c|c|}		
			\hline
			&$v_{1}^2/v_F^2$&$v_{2}^2/v_F^2$&$v_{12}^2/v_F^2$&$s^2/v_F^2$\\[0.6em] \hline
			$K_c$&$\dfrac{\gamma_c^2K_c^{-2}+1}{\gamma_c^2+1}$
			&$\dfrac{\gamma_c^2+K_c^{-2}}{\gamma_c^2+1}$
			&$\gamma_c\dfrac{K_c^{-2}-1}{\gamma_c^2+1}$
			&$\dfrac{K_c^{-2}+\gamma_c^2}{\gamma_c^2K_c^2+K_c^{-2}}$
			\\[0.6em] \hline
			$1/3$& 8.2&1.8&2.4&1.8\\
			0.2&22.6&3.4&7.2&1.3\\
			0.18&27.9&4.0&9.0&1.3\\
			0.1&90.1&10.9 &29.7&1.1\\\hline
		\end{tabular}	
		\caption{Effective velocities $v_{1}$, $v_2$, $v_{12}$ [see  Eq.~(\ref{eqn:vij})], and the velocity of the first mode $s$ [see Eq.~(\ref{eqn:s})] for different values of interaction parameter $K_c$. The inverse filling factor is fixed to $\gamma_c = 3$.}	
		\label{table:velocities}
	\end{table}

	\section{Refermionization at $K_c \approx 0.18 $ \label{sec: refermionization}}

We study the action given by~Eqs.~(\ref{eqn:action-gapped})--(\ref{eqn:action-coupling}), describing a quantum wire in the region $0<x<L$. We assume that the quantum wire is adiabatically connected to the leads at $x=0$ and at $x=L$. Although it is more standard to describe the leads as an extension of the wire to $x<0$ and $x>L$ with space-dependent LL interaction parameter $K_c$ \cite{MaslovStonePRB1995}, in this section we adopt an alternative approach introduced by Egger and Grabert~\cite{EggerGrabertPRB1998}. In this formalism, the coupling  to the leads enters via the boundary conditions for expectation values of density and current operators: 
\begin{align}
	&\frac{v_F}{K_c^2} \left \langle {\rho}_c \right\rangle + \left \langle j_c \right\rangle = \frac{2}{\pi}\int d\varepsilon\;  n_{L} (\varepsilon - V/2 ),& x=0,\label{eq:Egger-Grabert-charge-L}\\	
	&\frac{v_F}{K_c^2} \left \langle {\rho}_c \right\rangle - \left \langle  j_c \right\rangle = \frac{2}{\pi}\int d\varepsilon\;  n_{R} (\varepsilon + V/2 ),& x=L,\label{eq:Egger-Grabert-charge-R}\\
	&v_F \left \langle {\rho}_\sigma \right\rangle + \left \langle  j_\sigma \right\rangle = 0,& x=0,\\
	&v_F \left \langle {\rho}_\sigma \right\rangle - \left \langle  j_\sigma \right\rangle = 0,& x=L,\label{eq:Egger-Grabert-spin-R}
 	\end{align} 
	where $\rho_c = -\sqrt{2}\partial_x \phi_c / \pi$, $\rho_\sigma = \sqrt{2} \partial_t \theta_\sigma/(\pi v_F)$ are charge and spin densities, respectively, and $j_c = \sqrt{2}\partial_t \phi_c/\pi$, $j_\sigma = -\sqrt{2}v_F\partial_x \theta_\sigma/\pi$ are charge and spin currents, respectively\cite{footnote1}, $n_L$ and $n_R$ are electron distribution functions in the left and right leads. We note that the factor $K_c^{-2}$ in~Eqs.~(\ref{eq:Egger-Grabert-charge-L})--(\ref{eq:Egger-Grabert-charge-R}) takes into account a local potential drop between the quantum wire and the screening backgate at the contacts caused by electrons injected from the leads~\cite{EggerGrabertPRB1998}.
	
The action $S_E$ [see Eqs.~(\ref{eqn:action-gapped})--(\ref{eqn:action-coupling})] and corresponding boundary conditions can  in principle be refermionized at some special value of the LL interaction parameter $ K_c^*$, at the so-called Luther-Emery point~\cite{GiamarchiBook}. As a result, in terms of the new fermionic variables, the quadratic cross-term $S_{12}$ given by  Eq.~(\ref{eqn:action-coupling}) will be transformed into a non-linear term, consisting of four fermionic operators, and it will be still complicated to tackle this problem. Thus, prior to refermionization of the actions we perform the following shift of the bosonic field $\phi_2$:	
	\begin{align}
	\phi_2 \to \phi_2 + \frac{v_{12}^2}{v_2^2}\phi_1,
	\end{align}	
	transforming the density-density coupling $\partial_x \phi_1 \partial_x \phi_2$ into a current-current coupling $\partial_{\tau}\phi_1 \partial_{\tau} \phi_2$, which vanishes in the static limit. The action $S_E$ [see ~Eqs.~(\ref{eqn:action-gapped})--(\ref{eqn:action-coupling})] becomes
	\begin{align}
	\begin{split}
	&S_1 =\; -\frac{1}{\pi v_F}\int dxd\tau\; \left\{ \frac{\left(\partial_\tau \phi_1 \right)^2}{2}\left(1 + \frac{v_{12}^4}{v_2^4} \right) \right.\\&\left.+ \left(v_1^2 - \frac{v_{12}^4}{v_{2}^2} \right) \frac{\left(\partial_x \phi_1\right)^2}{2}+ \omega_0^2\cos \left(\sqrt{2}\lambda\phi_1\right)\right\} ,
	\end{split}\label{eqn:action-mode1-shifted}\\
	&S_2 =\; -\frac{1}{\pi v_F}\int dxd\tau\;\left\{ \frac{\left(\partial_\tau \phi_2 \right)^2}{2} + \frac{v_{2}^2 }{2} \frac{\left(\partial_x \phi_2 \right)^2}{2}\right\},\\
	&S_{12} =\;-\frac{1}{\pi v_F}\int dxd\tau\; \frac{v_{12}^2}{v_2^2} \partial_\tau \phi_1 \partial_\tau \phi_2.
	\end{align}

Next, we perform the refermionization by introducing new bosonic variables $\tilde{\varphi}_{1,2}$ and $\tilde{\theta}_{1,2}$ as well as new fermionic operators $R_{1,2}$ and $L_{1,2}$ as follows:
\begin{align}
	&\tilde{\varphi}_1 = \Xi \phi_1,\quad
	\tilde{\varphi}_2 = \sqrt{\frac{v_{2}}{{v_F}}} \phi_2,\\
	&\tilde{\theta}_1 = \Xi^{-1}\theta_1,\quad
	\tilde{\theta}_2 = \sqrt{\frac{v_F}{v_{2}}}\theta_2,\\
	&R_j \sim e^{-i\tilde{\varphi}_j + i\tilde{\theta}_j}, \quad L_j \sim e^{i\tilde{\varphi}_j +i\tilde{\theta}_j}.
	\end{align}
We note that only if the cosine term in $S_1$ is of the form $\cos (2\tilde{\phi}_1)$ in terms of the rescaled bosonic field $\tilde{\phi}$ [see Eq.~(\ref{eqn:action-mode1-shifted})], it converts into a simple quadratic term
	$R_1^\dag L_1 + L_1^\dag R_1$. For this to be the case, the following condition must be fulfilled,
\begin{align}
&\sqrt{2}\lambda \Xi^{-1} = 2,
\label{eqn:Luther-Emery-condition}\\
	&\Xi =  \left[\frac{v_1^2}{v_F^2} \left(1 - \frac{v_{12}^4}{v_1^2 v_2^2} \right)\left( 1+ \frac{v_{12}^2}{v_2^2}\right)\right]^{\frac{1}{4}}.
\end{align}
In case of the effective three-particle scattering shown in Fig.~\ref{fig:spectrum},  corresponding to the filling factor $\nu=1/3$ with $\gamma_c = 3$, the condition defined in  Eq.~(\ref{eqn:Luther-Emery-condition}) yields the value of the interaction parameter, 
\begin{align}
	K_c^* = \frac{1}{3}\sqrt{\frac{4}{15}\sqrt{19} - \frac{13}{15}}\approx 0.18.
	\end{align}	 
The refermionized Hamiltonian then becomes $H=H_{1} + H_{2} + H_{12}$, with
	\begin{align}
	\begin{split}
	&H_1 = \int dx\; is \left[-R_1^\dag \partial_x R_1 + L_1^\dag \partial_x L_1\right] \\&\hspace{100pt}+ \Delta \left(R_1^\dag L_1 + L_1^\dag R_1 \right),
	\end{split}
	\\
	&H_2 = \int dx\; iv_2 \left[-R_2^\dag \partial_x R_2 + L_2^\dag \partial_x L_2\right],\\
	\begin{split}
	&H_{12} = \frac{\sqrt{v_1 v_2}v_{12}^2}{v_2^2} \int dx_1 dx_2\; \left[\rho_{R1}(x_1) - \rho_{L1}(x_1) \right]\\&\times\left[\rho_{R2}(x_1) - \rho_{L2}(x_1) \right],
	\end{split}\label{eqn:refermionization-crossterm}
	\end{align}		
where $\Delta$ is an energy gap, the operators $\rho_{R(L),j}$ are density operators of right (left)-moving ``refermions'' corresponding to the $j$-th mode, and $s$ is a velocity of the first mode defined as 
	\begin{align}
	s^2 = \frac{v_1^2 v_2^4 - v_{12}^4 v_2^2}{v_2^4 + v_{12}^4 }
	  = v_F^2\frac{K_c^{-2} + \gamma_c^2}{K^{-2}_c + \gamma_c^2 K_c^2}.
	\label{eqn:s}
	\end{align}	
From an RG study (see Appendix~\ref{app:RG}) we obtain that the cross-term $H_{12}$ is irrelevant in the low-energy limit, so we disregard it in this section. Later in Sec.~\ref{sec:semiclassical}, the effect of $H_{12}$, however, will be discussed 
	
After disregarding the cross-term the Hamiltonian becomes quadratic in terms of fermionic fields.  The  equations of motion (in real-time representation) for the fields $R_j$, $L_j$ read as
\begin{align}
	&i\partial_t R_1 = -i s \partial_x R_1 + \Delta(x) L_1,\label{eqn:refermion-motion-0}\\
	& i\partial_t L_1 = i s \partial_x L_1 + \Delta(x) R_1, 	\label{eqn:refermion-motion-1}\\
	&i\partial_t R_2 = -i v_2 \partial_x R_2,\, \ \ \  i\partial_t L_2 = i v_2 \partial_x L_2. 
	\label{eqn:refermion-motion-2}
	\end{align}
	Here we assume that in general  the partial gap may vary with the coordinate $x$. The density of ``refermions'' is given by $\rho_j = R_j^\dag R_j + L_j^\dag L_j$. From the continuity equation, $\partial_t \rho_j + \partial_x j_j = 0$, the currents of ``refermions'' can be defined as
	\begin{align}
	j_1 = s\left(R_1^\dag R_1 - L_1^\dag L_1 \right),\ j_2 = v_2\left(R_2^\dag R_2 - L_2^\dag L_2 \right).
	\end{align}
	
	The ``refermion'' densities and currents are related to physical density and current operators by the following expressions: 
	\begin{align}
 &\rho_c = \frac{\sqrt{2}}{\lambda} \left[\rho_1\left(\gamma_c - \frac{v_{12}^2}{v_2^2} \right) \Xi^{-1} - \rho_2\sqrt{\frac{v_F}{v_2}}\right],\label{eqn:refermion-relation1}\\
 &j_c = \frac{\sqrt{2}}{\lambda} \left[j_1\left(\gamma_c - \frac{v_{12}^2}{v_2^2} \right) \Xi^{-1} - j_2\sqrt{\frac{v_F}{v_2}}\right],\label{eqn:refermion-relation2}\\ 	
 &v_F \rho_\sigma = \frac{\sqrt{2}}{\lambda}\left[j_1\left(1 + \gamma_c \frac{v_{12}^2}{v_2^2} \right) \Xi^{-1} + j_2\gamma_c \sqrt{\frac{v_F}{v_2}}\right],\\ 
 &j_\sigma = \frac{\sqrt{2}v_F}{\lambda}\left[\rho_1\left(1 + \gamma_c \frac{v_{12}^2}{v_2^2} \right) \Xi^{-1} + \rho_2\gamma_c \sqrt{\frac{v_F}{v_2}}\right]\label{eqn:refermion-relation4}. 	
	\end{align}
	
	Now we solve Eqs.~(\ref{eqn:refermion-motion-0})--(\ref{eqn:refermion-motion-2}) with boundary conditions given by Eqs.~(\ref{eq:Egger-Grabert-charge-L})--(\ref{eq:Egger-Grabert-charge-R}) and with an additional boundary condition corresponding to adiabatically attached contacts,
	\begin{align}
	\left\langle R^\dag_1(x=0) L_1(x=L)  \right\rangle = \left\langle L^\dag_1(x=L) R_1(x=0)  \right\rangle = 0,
	\label{eqn:refermion-boundary}
	\end{align}
which means that the right-movers injected from the left lead are independent from the left-movers injected from the right lead.

	\subsection{Zero-temperature conductance for fine-tuned value of chemical potential \label{sec:refermionization-T0}}
	
	First, we consider the zero-temperature limit, recovering known results for the conductance of a fractional Luttinger liquid.	In this section we also assume for  simplicity that the gap abruptly vanishes in the leads $\Delta(x) = \Delta \Theta(x)\Theta(L-x)$. First, from Eq.~(\ref{eqn:refermion-motion-2}), we note that current and density of the refermions corresponding to the second mode do not depend on the coordinate, {\it i.e.}, $\rho_2(x) = \rho_2$, $j_2(x) = j_2$. Furthermore, if the voltage bias is applied symmetrically, $\rho_i(x=0) = -\rho_i(x=L)$, and, hence, $\rho_2 = 0$. In adittion, from Eqs.~(\ref{eqn:refermion-motion-0})--(\ref{eqn:refermion-motion-1}), it is easy to obtain a general form of the solution in energy representation for $|\varepsilon|<\Delta$,  
	\begin{multline}
	\begin{pmatrix}
	R_1(x,\varepsilon)\\
	L_1(x,\varepsilon)
	\end{pmatrix}
	= \frac{A(\varepsilon)}{\sqrt{2} \Delta}
	\begin{pmatrix}
	\Delta\\
	\varepsilon - i\kappa s
	\end{pmatrix} e^{-\kappa x} \\+
	\frac{B(\varepsilon)}{\sqrt{2}\Delta}
	\begin{pmatrix}
	\Delta\\
	\varepsilon + i\kappa s
	\end{pmatrix} e^{\kappa (x-L) },
	\label{eqn:solution-below-gap}
	\end{multline}
where $\kappa = \sqrt{\Delta^2 - \varepsilon^2}/s$ and $A(\varepsilon)$ as well as $B(\varepsilon)$ are energy-dependent fermionic operators corresponding to decaying waves propagating from the left and the right leads. Thus, the tunneling current carried by gapped refermions is given by
	\begin{align}
	j_1(\varepsilon) = -is^2 \kappa e^{-\kappa L}\left[ A^\dag B/\left(\varepsilon - i\kappa s \right) - H.c.  \right],
	\label{eqn:refermion-tunneling-current}
	\end{align}
	and is exponentially small in long wires, $L \gg s/\Delta$ (see Appendix~\ref{app:refermionization} for details). Neglecting the tunneling current $j_1$ and making use of Eqs.~(\ref{eqn:refermion-relation1})--(\ref{eqn:refermion-relation4}), we arrive at the following relations between charge and spin densities/currents:
\begin{align}
	v_F\rho_\sigma + \gamma_c j_c = 0,\ \ \ \ 
	 	 v_F\rho_c = \gamma_c K_c^2 j_\sigma.
	\end{align}
These relations along with the boundary conditions given by Eqs.~(\ref{eq:Egger-Grabert-charge-L})--(\ref{eq:Egger-Grabert-spin-R}) can be treated as a system of linear equations to be solved in order  to obtain coordinate-independent $\rho_\sigma$, $j_c$, and coordinate-dependent $\rho_c$, $j_\sigma$, {\it i.e. }$\rho_c(x=0)$, $\rho_c(x=L)$, $j_\sigma(x=0)$, $j_\sigma(x=L)$. Finally, the charge current can be related to the applied bias voltage $V$ as
\begin{align}
j_c = \frac{V}{\pi\left(\gamma_c^2 + 1 \right)}.
\end{align}
Restoring dimensional units,  we obtain the zero-temperature limit of the conductance which is in agreement with previous results\cite{OregPRB2014, MengPRB2014}: 
\begin{align}
G_\nu =  j_c/V=\frac{2e^2}{h} \frac{1}{\gamma_c^2 + 1}.
\label{eqn:fractional}
\end{align}

The tunneling contribution can be found by expressing expectation values of densities and currents in terms of the thermodynamic averages: $\langle A^\dag A \rangle$, $\langle B^\dag B \rangle$, $\langle A^\dag B \rangle$, $\langle B^\dag A \rangle$ and imposing the boundary conditions given by Eqs.~(\ref{eq:Egger-Grabert-charge-L})--(\ref{eq:Egger-Grabert-spin-R}) and Eq.~(\ref{eqn:refermion-boundary}) (see Appendix~\ref{app:refermionization} for details). The cross-correlators $\langle A^\dag  B  \rangle$ proportional to the tunneling transparency of the effective barrier created by the gap are exponentially small, $\langle A^\dag (\varepsilon) B (\varepsilon) \rangle, \langle B^\dag(\varepsilon) A(\varepsilon) \rangle \propto e^{-\kappa L}$. An extra exponential factor arises from~Eq.~(\ref{eqn:refermion-tunneling-current}), and therefore tunneling contributions to the conductance at $K_c = K_c^*$ in the limit $L\to\infty$ can be estimated as
\begin{align}
\delta G \sim G_0 e^{-2 \Delta L /s}.
\label{eqn:delta-G}
\end{align}
The resulting dependence of the conductance on the wire length $L$ at $T=0$ is shown in~Fig.~\ref{fig:conductance-T0} (see details of numerical calculations in~Appendix~\ref{app:refermionization}). As the tunneling through a gap in short wires becomes significant,  $L \lesssim s/\Delta \sim v_F/\Delta$, the conductance differs from the fractional value given by Eq.~($\ref{eqn:fractional}$). We note that in this limit the correlation length in~Eq.~(\ref{eqn:1st-RG}) is determined by the length of the wire, $l_c \sim L$.
	
		\begin{figure}
		\includegraphics[width=\columnwidth]{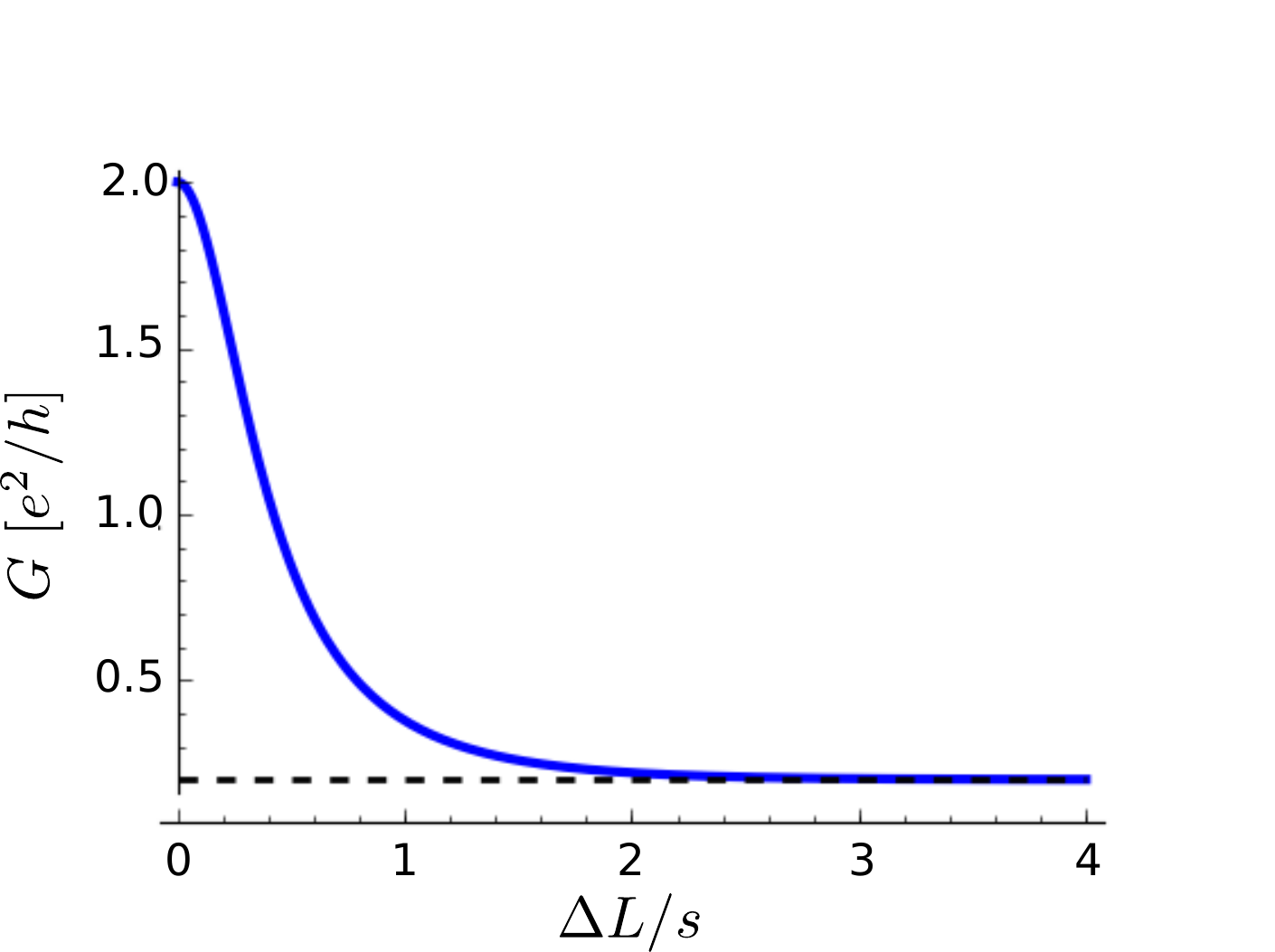}
		\caption{Dependence of conductance $G$ at the filling factor $\nu = \gamma_c^{-1} = 1/3$ on length $L$ at zero temperature ($T=0$) for the interaction parameter $ K_c^* \approx 0.18$. 		The value of the gap $\Delta$ renormalized by interactions is defined by Eqs.~(\ref{eqn:1st-RG})--(\ref{eqn:1st-RG-c}). The blue line shows the result obtained numerically~(see Appendix~\ref{app:refermionization}) when the tunneling contribution defined in Eq.~(\ref{eqn:refermion-tunneling-current}) is taken into account; the dashed line shows the fractional conductance $G_{\nu=1/3} = e^2/5h$. 
The tunneling contribution to the conductance vanishes exponentially for long wires ($L\gg \hbar s/\Delta$), $G \approx G_{\nu=1/3} + G_0e^{-2\Delta L/s}$. The obtained conductance differs from the fractional value $G_{\nu=1/3}$ in short wires $L \lesssim \hbar s/\Delta \sim \hbar v_F/\Delta$ as the tunneling through a gap becomes significant. In the limit of a short wire, $L\ll \hbar s/\Delta$, the fractional conductance is no longer observed, $G = 2 G_0=2e^2/h$. Note that in this limit the correlation length in~Eq.~(\ref{eqn:1st-RG}) is determined by the length of the wire, $l_c \sim L$, and the gap is given by Eq.~(\ref{eqn:1st-RG-c}).}
		\label{fig:conductance-T0}
	\end{figure}	
	
	\subsection{Finite-temperature conductance for fine-tuned value of chemical potential \label{sec:refermionization-T}}
	
		\begin{figure}
		\includegraphics[width=\columnwidth]{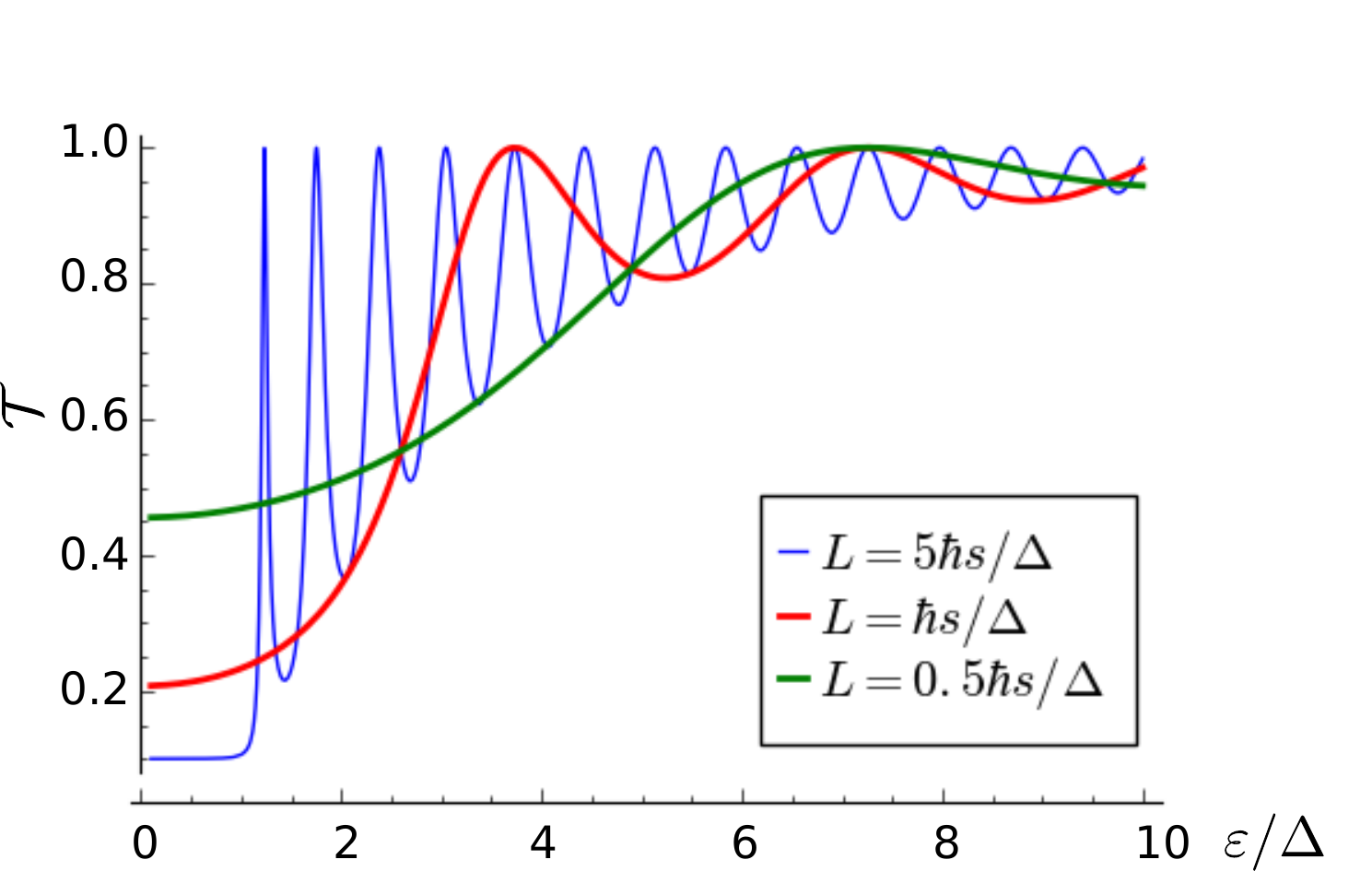}
		\caption{Effective transmission probability $\mathcal{T}$ [see~Eq.~(\ref{eqn:transmission})] for a long wire (blue curve) and short wires (red and green curves). Calculations have been performed numerically~(see Appendix~\ref{app:refermionization}) for an abrupt coordinate dependence of the gap, $\Delta(x)=\Delta\Theta(x)\Theta(L-x)$. At energies less than the gap $\Delta$, the effective transmission is reduced due to a scattering at the effective potential barrier. At higher energies, Fabry-Perot oscillations emerge with a period $\delta \varepsilon \sim hs /2L$. The transmission becomes ideal, $\mathcal{T}=1$, in the limit $\varepsilon \gg \Delta$. Note, that for the short wire (green curve) the gap has been calculated using~Eq.~(\ref{eqn:1st-RG-c}).}
		\label{fig:transmission}
	\end{figure}
	
	Now we can proceed with the more general case of non-zero temperature $T>0$. In energy representation it is convenient to define density and current operators $\rho_{1,2}(\varepsilon)$, $j_{1,2}(\varepsilon)$ as
	\begin{align}
	&\rho_j(\varepsilon) = R^\dag_j(\varepsilon)R_j(\varepsilon) + L^\dag_j(\varepsilon)L_j(\varepsilon),\\
	&j_1(\varepsilon) = s\left[R_1^\dag(\varepsilon) R_1(\varepsilon) -L_1^\dag(\varepsilon)L_1(\varepsilon)\right],\\
	&j_2(\varepsilon) = v_2\left[R_2^\dag(\varepsilon) R_2(\varepsilon) -L_2^\dag(\varepsilon)L_2(\varepsilon)\right],
	\end{align}
so that physical charge/spin density and current operators $\rho_\nu(\varepsilon)$, $j_\nu(\varepsilon)$ can be defined by linear relations, see Eqs.~(\ref{eqn:refermion-relation1})--(\ref{eqn:refermion-relation4}).
Since the boundary conditions [see~Eqs.~(\ref{eq:Egger-Grabert-charge-L})--(\ref{eq:Egger-Grabert-spin-R})] and the equations of motion [see Eqs.~(\ref{eqn:refermion-motion-1})--(\ref{eqn:refermion-motion-2})] are linear, the charge current can be linearly related to the difference of Fermi distribution functions in the left and right leads,
	\begin{align}
	j_c(\varepsilon) = 2\mathcal{T}(\varepsilon)\left[n_L\left(\varepsilon - \frac{V}{2}\right) - n_R\left(\varepsilon + \frac{V}{2}\right) \right],	
	\label{eqn:transmission}
	\end{align}
	where the proportionality coefficient $\mathcal{T}(\varepsilon)$ can be interpreted as an effective transmission probability of the quantum wire (for details see Appendix~\ref{app:refermionization}). Integrating Eq.~(\ref{eqn:transmission}) over energy, we arrive at the generalized
	 Landauer formula~\cite{ButtikerPRB1985, Datta1997},
	\begin{align}
	G = 2G_0 \int\limits_{-\infty}^{+\infty} d\varepsilon\; \mathcal{T}(\varepsilon) \frac{1}{4T \cosh^2\left(\varepsilon/2T\right)}.
	\label{eqn:Landauer}
	\end{align}
In the limiting case, when the gap $\Delta(x)$ varies adiabatically from zero at the contact to some finite value $\Delta$ inside the wire and, in addition, when the wire is long $L\gg s/\Delta$, the effective transmission probability is simply given by~(see~Appendix~\ref{app:refermionization} for details),
	\begin{align}
	\mathcal{T}(\varepsilon) = \Theta(|\varepsilon| - \Delta) + \frac{1}{\gamma_c^2+1} \Theta(\Delta-|\varepsilon|).
	\end{align}
	The first term describes an ideal transmission at energies above the gap. The second term responsible for the fractional conductance can be derived similarly as in the previous section. The finite-temperature conductance in this adiabatic limit is given by a simple expression,
	\begin{align}
	G = \frac{2e^2}{h\left(\gamma_c^2 +1 \right)}\left(1 + 2\gamma_c^2\frac{e^{-\Delta/T}}{1+ e^{-\Delta/T}}\right).
	\label{eqn:G-T-dependence}
	\end{align}
	Note that the chemical potential is inside the gap, and the difference from fractional value $G_{\nu}$ is caused by thermal electrons with energies above the gap propagating through the wire.
	
\begin{figure}
		\includegraphics[width=\columnwidth]{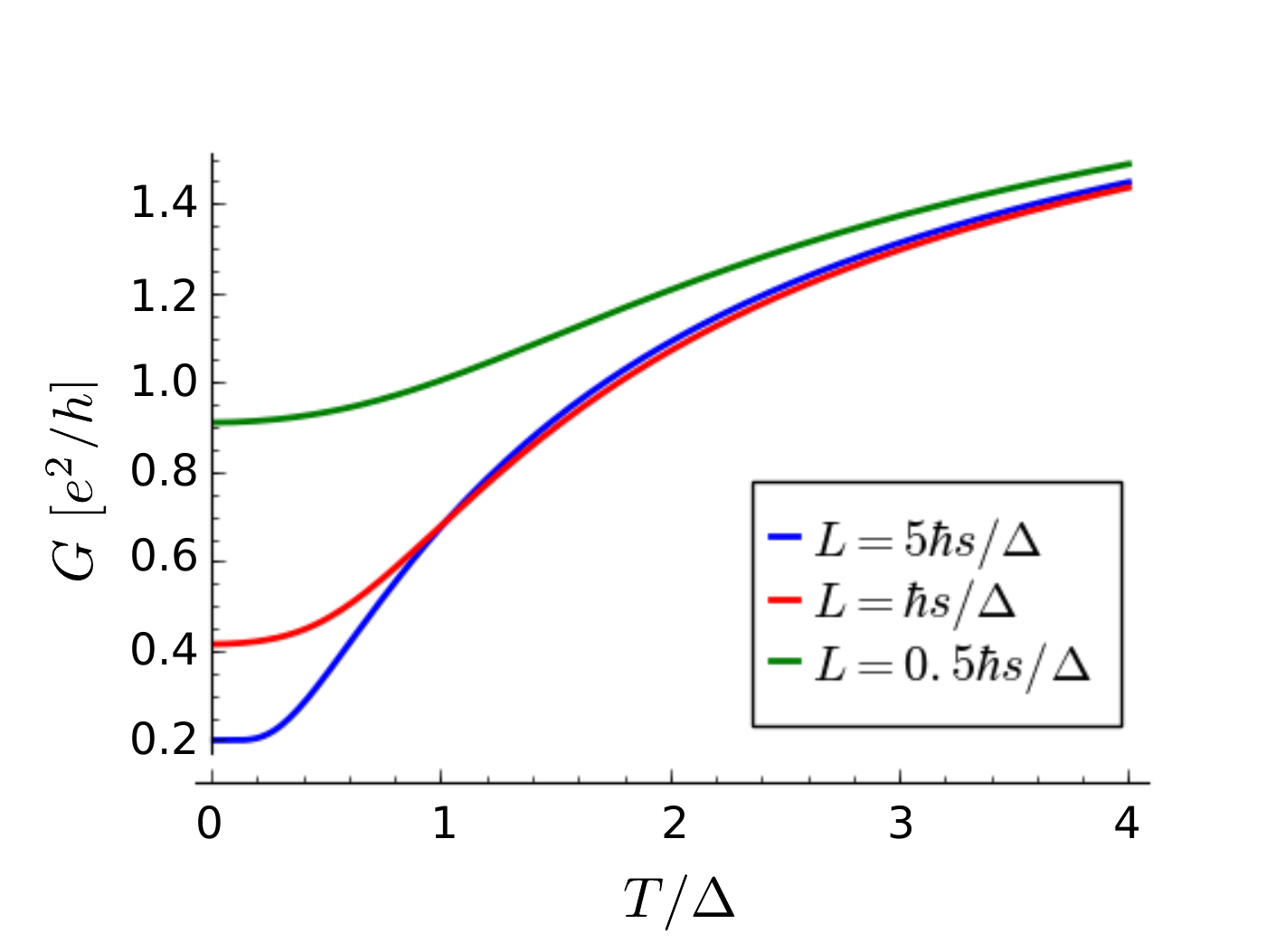}
		\caption{Conductance $G$ of a quantum wire in the regime of fractional helical liquid, $\nu=1/3$, as function of temperature for different lengths of the wire: $L = 5\hbar s/\Delta$ (blue curve), $L = \hbar s/\Delta$ (red curve), $L = 0.5\hbar s/ \Delta$ (green curve). Results are obtained numerically [see Eq.~(\ref{eqn:Landauer})] by using with effective transmission probability $\mathcal{T}$ from Fig. \ref{fig:transmission}.
Even at temperatures $T \gtrsim \Delta$, the conductance does not reached its full value  $2e^2/h$. We note that at high temperatures, $G$ only weakly depends on the length $L$. 
	}	
\label{fig:conductance-T}
	\end{figure}

In the opposite limiting case, in which the gap drops abruptly in the leads, $\Delta(x) = \Delta\Theta(x)\Theta(L-x)$, the effective transmission probability is a more complicated fucntion, manifesting itself in Fabry-Perot oscillations (see~Fig.~\ref{fig:transmission}).	Using Eq.~(\ref{eqn:Landauer}), we obtain  the temperature-dependence of  the conductance~(see Fig.~\ref{fig:conductance-T})  that shows activation behavior. However, even at temperatures $T \gtrsim \Delta$, the conductance does not reach its full value  $2e^2/h$ due to the presence of the gap.

			\begin{figure}[b]
		\includegraphics[width=\columnwidth]{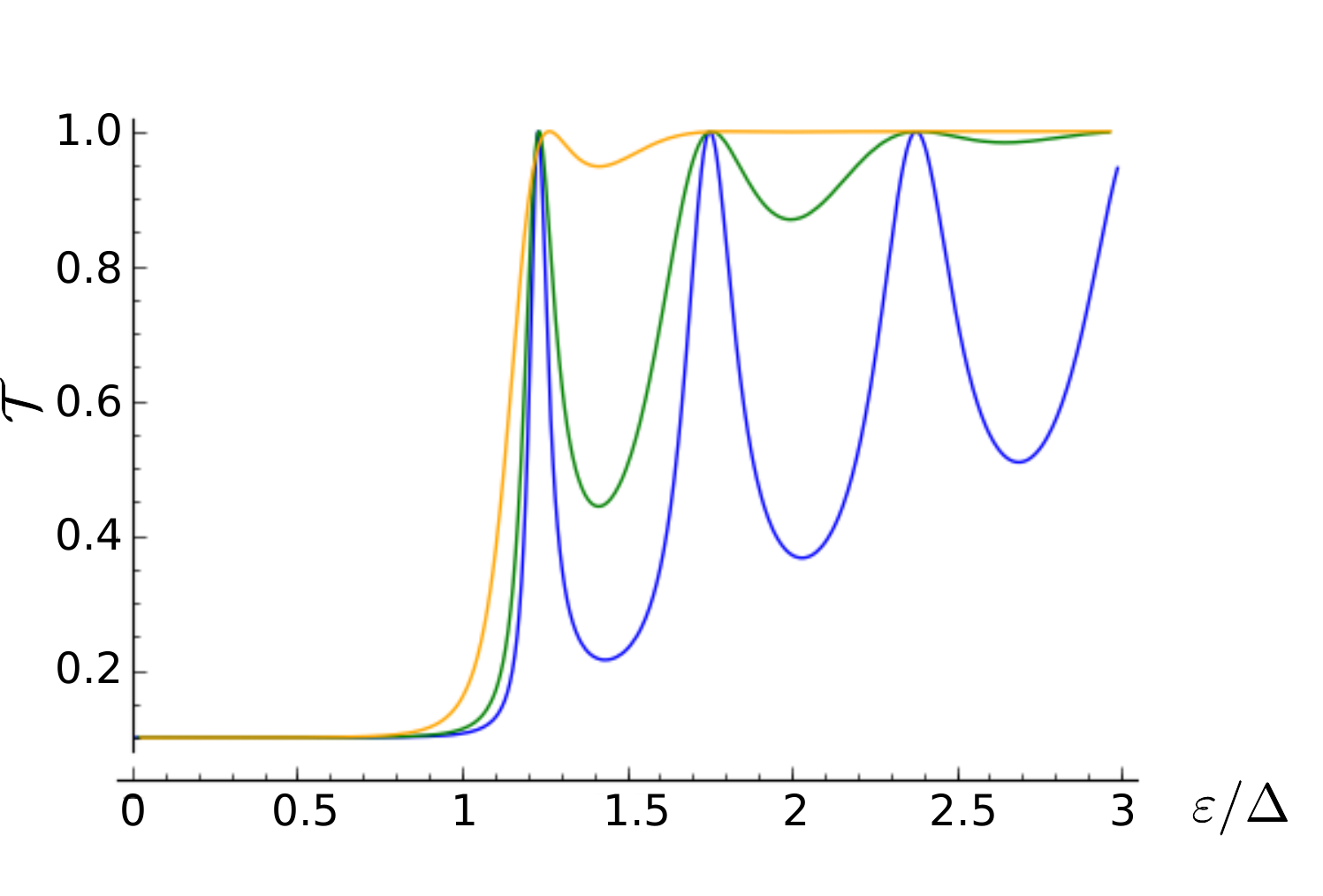}
		\caption{Effective transmission probability $\mathcal{T}$ for different profiles of partial gap $\Delta(x)$ [see Eq.~(\ref{eqn:gap-profile})]: $l\ll l_\Delta = \hbar s/\Delta$ (blue curve), $l=0.5 l_\Delta$ (green curve) and $l=l_\Delta$ (orange curve).  In case of a smooth gap profile, the Fabry-Perot oscillations are washed out already at $\varepsilon \sim \Delta$ while, in case of an abruptly changing gap, they vanish only at $\varepsilon \gg \Delta$. The length of the wire is fixed to $L= 5 \hbar s/ \Delta$.}	
		\label{fig:transmissions}
	\end{figure}

\begin{figure}[!t]
	\includegraphics[width=\columnwidth]{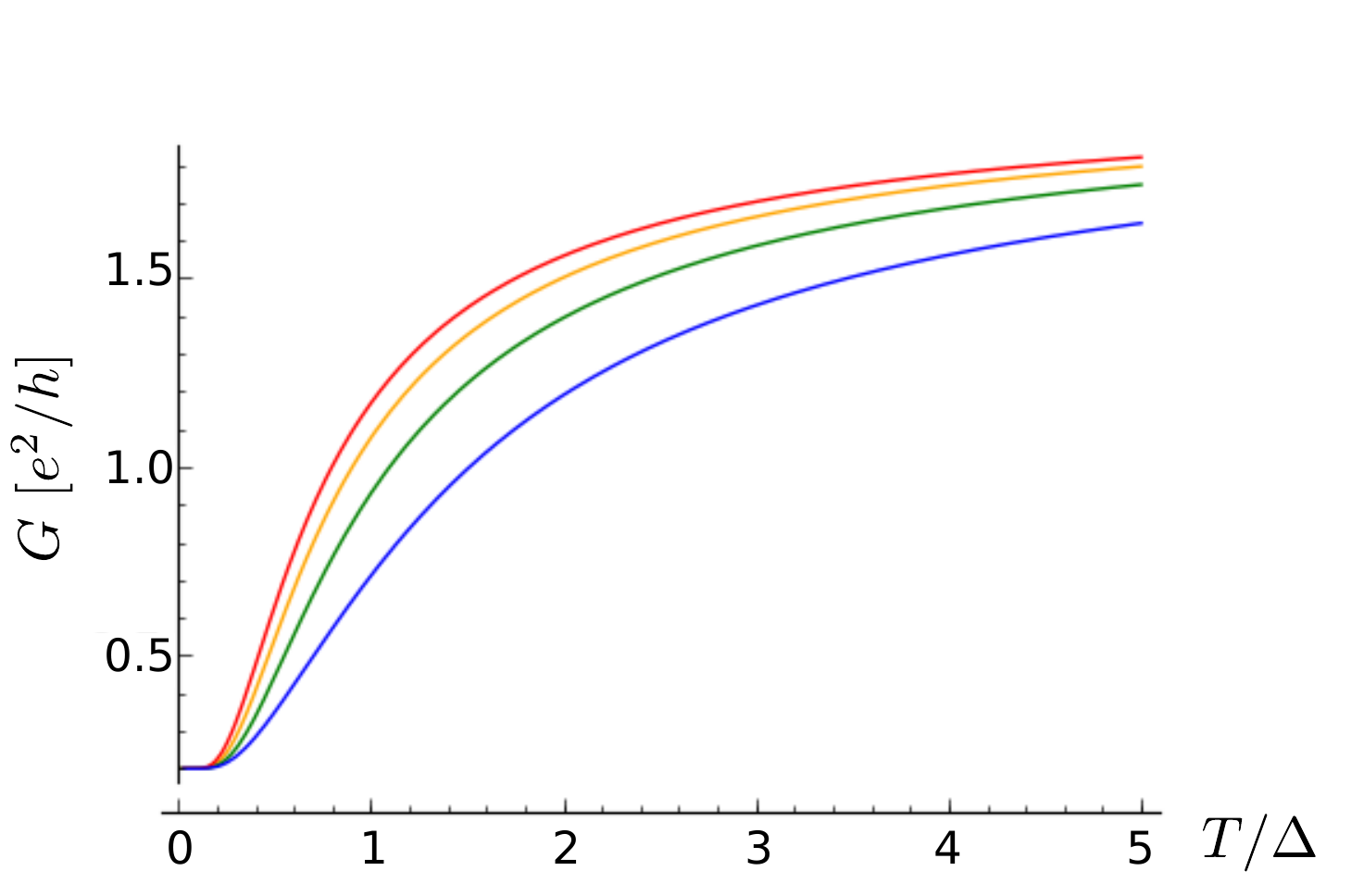}
	\caption{Conductance $G$ of a quantum wire of length  $L=20\hbar v_F/L$ in the regime of fractional helical liquid, $\nu=1/3$, as function of temperature for different profiles of the gap given by~Eq.~(\ref{eqn:gap-profile}):  $l \gg \hbar v_F/\Delta$ (red line), $l = \hbar v_F/\Delta$ (orange line), $l = 0.5\hbar v_F/\Delta$ (green line), $l \to 0$ (blue line).
	 The results were obtained using Eq.~(\ref{eqn:Landauer}) with effective transmission probability calculated numerically (see Fig.~\ref{fig:transmissions})
	 The dependence shows a steeper activation behavior as the gap profile becomes smoother.	}
	\label{fig:conductance-profiles}
\end{figure}

We also study an intermediate case, assuming that the gap $\Delta(x)$ is modeled by a smooth profile along the wire of the form given by Eq.~(\ref{eqn:gap-profile}). The resulting transmissions are shown in Fig.~\ref{fig:transmissions}. The Fabry-Perot oscillations at $\varepsilon \gtrsim \Delta$ are washed out if the characteristic length  $l$ at which gap $\Delta(x)$ goes to zero is much larger than the lengthscale $l_\Delta = \hbar s/\Delta$ set by the partial gap. The resulting temperature dependence of the conductance for a long wire and different gap profiles is shown in Fig.~\ref{fig:conductance-profiles}.  If the modes in the leads and inside the wire are not well coupled, which corresponds to an abrupt change in the gap, the conductance is suppressed even at $T \gtrsim \Delta$, compared with the smooth gap profile.

\subsection{Dependence of conductance on chemical potential}

\begin{figure*}
	\includegraphics[width=\linewidth]{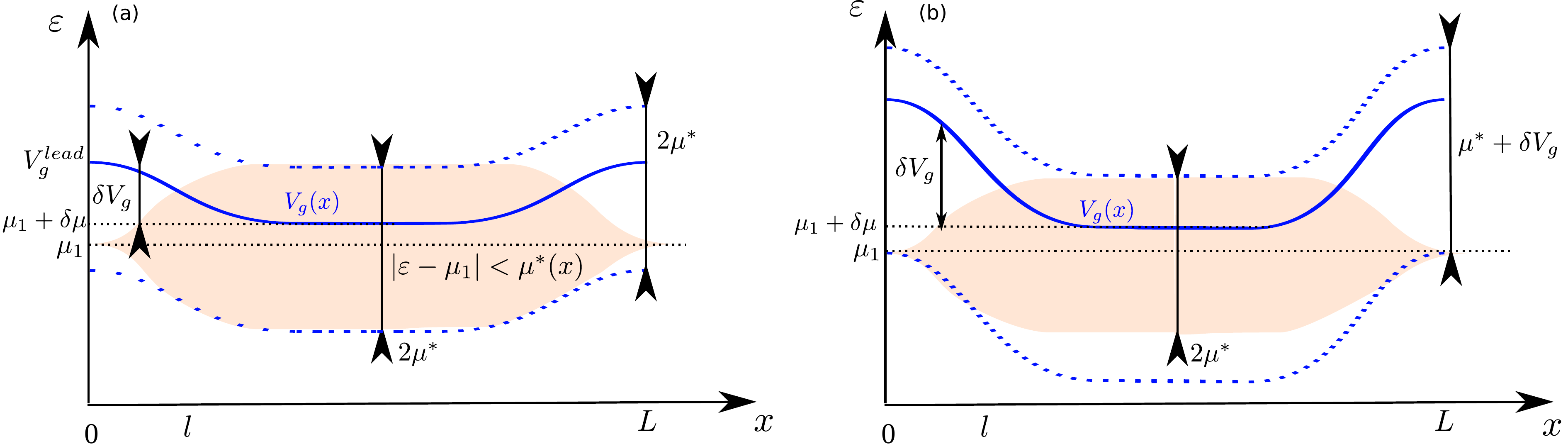}
	\caption{ Schematic energy diagram for a smooth gate potential drop at the contact in two limiting cases:
(a) small gate potential variation, $\delta V_g < \mu^*$; (b) large gate potential variation, $\delta V_g > \mu^*$. If the gate potential defined by Eq.~(\ref{eqn:gate-voltage}) [solid blue line] lies close to the resonance value of the chemical potential, \textit{i.e.} within the range of $\mu_1 - \mu^* < V_g < \mu_1 + \mu^*$ (shaded pink region), the measured conductance takes fractional value, $G_{\nu=1/3} = e^2/5h$.  The dotted blue lines show the corresponding bounds on the gate potential.
In panel (a), the range of values of $V_g^{lead}$ at which quantized values of conductance $G_{\nu}$ can be observed is given by $2\mu^*$.
In contrast to that, in panel (b), this range is broader and can be estimated as $\mu^*+\delta V_g$. }
	\label{fig:Landau-Zener}
\end{figure*}	

		\begin{figure}[b]
		\includegraphics[width=\columnwidth]{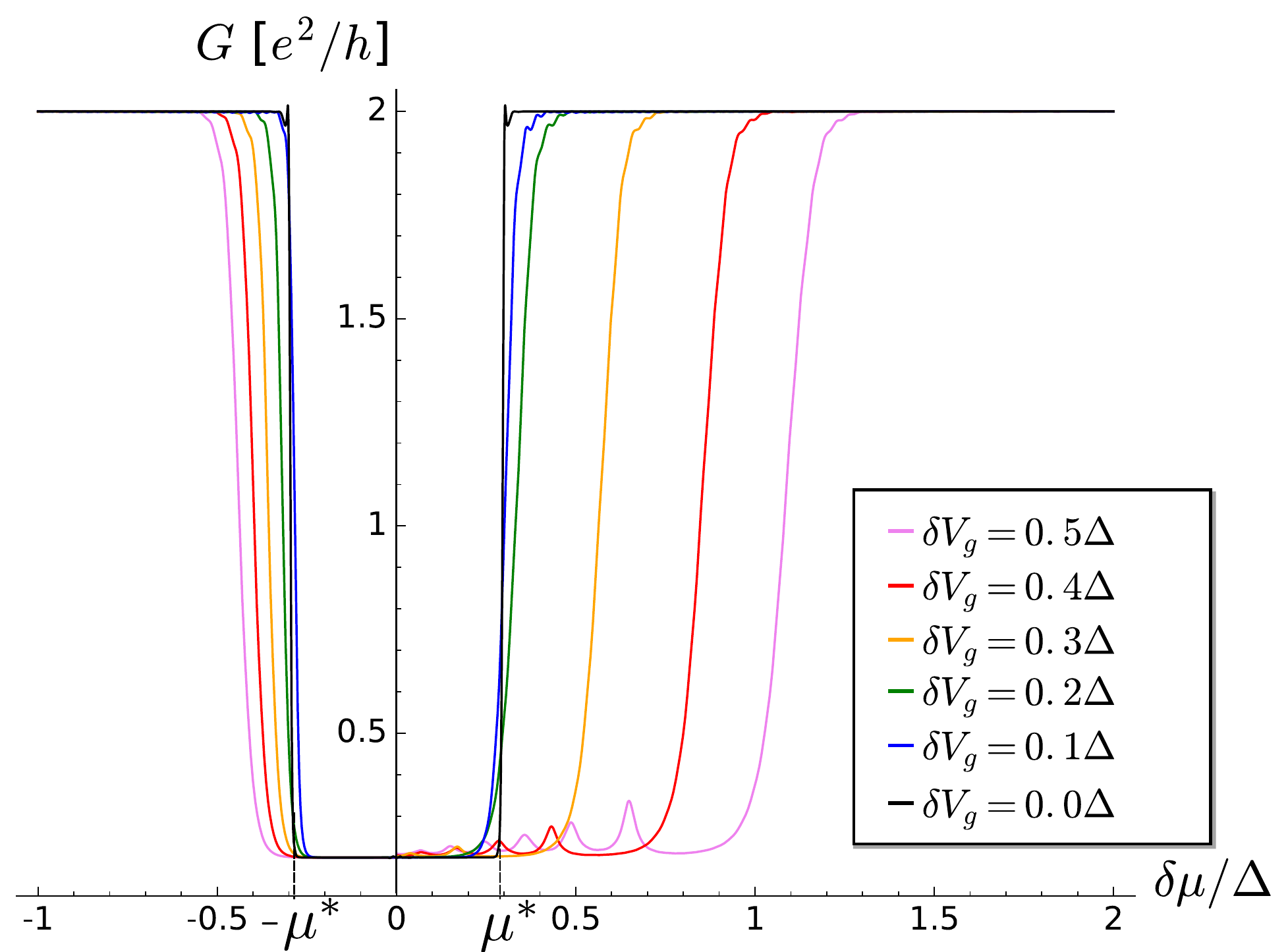}
		\caption{Dependence of conductance $G$ on the shift of chemical potential $\delta \mu$ for different values of gate potential variation $\delta V_g$ at the contact at zero temperature $T=0$. The partial gap and gate potential profiles are defined by Eqs.~(\ref{eqn:gap-profile}) and (\ref{eqn:gate-voltage}) with $l=l_V=5\hbar s/\Delta$ and the length of the wire is fixed to $L=20 \hbar s/\Delta$. The shift of chemical potential $\delta \mu$  measures the difference between the gate potential $V_g(x)$ in the bulk of the wire and at the resonance value of chemical potential $\mu_1$, see Fig.~\ref{fig:Landau-Zener}. At high values of gate voltage variation, $\delta V_g \gtrsim \mu^*\approx 0.3\Delta$, the dip in conductance caused by opening of the partial gap broadens, see also Fig.~\ref{fig:Landau-Zener}.}
\label{fig:G-vs-mu-adiabatic}
	\end{figure}
	
Next, we consider parameter regimes in which the chemical potential $\mu = \mu_n +\delta \mu$ is shifted away from the resonant value $\mu_n$, corresponding to the filling factor $\nu = k_F/k_{so} = 1/(2n+1)$. We focus on the case when the shift of the chemical potential is not significantly greater than the gap $\Delta$ obtained previously, and, therefore, assume that this shift is much less than the Fermi energy, $\delta \mu \ll \mu_n$. We linearize the fermionic fields $\psi_{s}(x)$ near the new  Fermi momenta.  The effective back-scattering interaction term resulting in the partial gap [see Eq.~(\ref{eqn:interaction-term})] is now replaced by
\begin{align}
\mathcal{O}_B = g_B^{(n)}\left(L^\dag_{\uparrow} R_{\uparrow} \right)^n R_{\uparrow}L_{\downarrow}^\dag \left(L_{\downarrow}^\dag R_{\downarrow} \right)^n e^{i(4n +2) x \delta \mu /v_F} + H.c.
\end{align} 
As a result, the sine-Gordon term in Eq.~(\ref{eqn:Matsubura-rho-sigma-action}) also acquires an additional position dependent phase $(4n+2) x \delta\mu/ v_F$. Without loss of generality, in what follows, we again focus on $n=1$ (corresponding to the filling factor $\nu=1/3$). As a consequence, the equations of motion involving fields $R_1$ and $L_1$~[see Eqs.~(\ref{eqn:refermion-motion-0})--(\ref{eqn:refermion-motion-1})] are replaced by 
	\begin{align}
	&i\partial_t R_1 = -i s \partial_x R_1 + e^{iKx}\Delta(x) L_1,
	\label{eqn:refermion-away1}
	\\ &i\partial_t L_1 = i s \partial_x  L_1 + e^{-iKx}\Delta(x) R_1,
		\label{eqn:refermion-away2}
	\end{align}
with the momentum shift $K=6 \delta\mu/ v_F$. We note that these equations can be brought back to the form of Eqs.~(\ref{eqn:refermion-motion-0})--(\ref{eqn:refermion-motion-1}) by a gauge transformation,
\begin{align}
	R_1(x,t) &\to\; R_1(x,t) e^{iK (x-st)/2},\\ 
	L_1(x,t) &\to\; L_1(x,t) e^{-iK (x+st)/2}.
\end{align}
It is also convenient to rewrite this transformation in energy representation,
\begin{align}
	R_1(x,\varepsilon) &\to\; R_1(x,\varepsilon + Ks/2) e^{iK x/2},\\ 
	L_1(x,\varepsilon) &\to\; L_1(x,\varepsilon + Ks/2) e^{-iK x/2}.
\end{align}
Thus, the effective transmission at the Fermi level $\mu = \mu_n + \delta \mu$ is the same as the effective transmission obtained in Sec.~\ref{sec:refermionization-T} [see Eq.~(\ref{eqn:transmission})] at $\varepsilon =  Ks/2 \equiv 3 s\delta \mu /v_F $. More generally, $\mathcal{T}_{\mu_n + \delta\mu}(\varepsilon) = \mathcal{T}_{\mu_n}(\varepsilon + Ks/2)$.
	
As a result, the conductance at zero temperature can be related to the effective transmission obtained earlier as
\begin{align}
	G_{\mu_1 + \delta\mu} = \frac{2e^2}{h}\mathcal{T}_{\mu_1}\left(\frac{3  s}{v_F} \delta \mu  \right).
	\end{align}
Thus, the fractional conductance $G_{\nu=1/3} = e^2/5h$ can be observed if the shift of the chemical potential is small enough, $|\delta \mu| < \mu^* = v_F \Delta / 3s \approx 0.3 \Delta$. The numerical value for the effective velocity $s$ was taken at $K_c = K_c^* \approx 0.18$ (see~Table~\ref{table:velocities}).   We note that, in contrast to the case of the standard Zeeman gap observed at $n=0$, the maximum value of the shift of chemical potential at which one can still observe fractional conductance values is less than the gap, $\mu^* < \Delta$. This is due to the fact that the higher-order interaction term  given by~Eq.(\ref{eqn:interaction-term}) results in a larger momentum mismatch if the chemical potential is away from the resonance value. Thus, a more precise tuning of the chemical potential is required for the observation of fractional conductances.

	\begin{figure}[b]
	\includegraphics[width=\columnwidth]{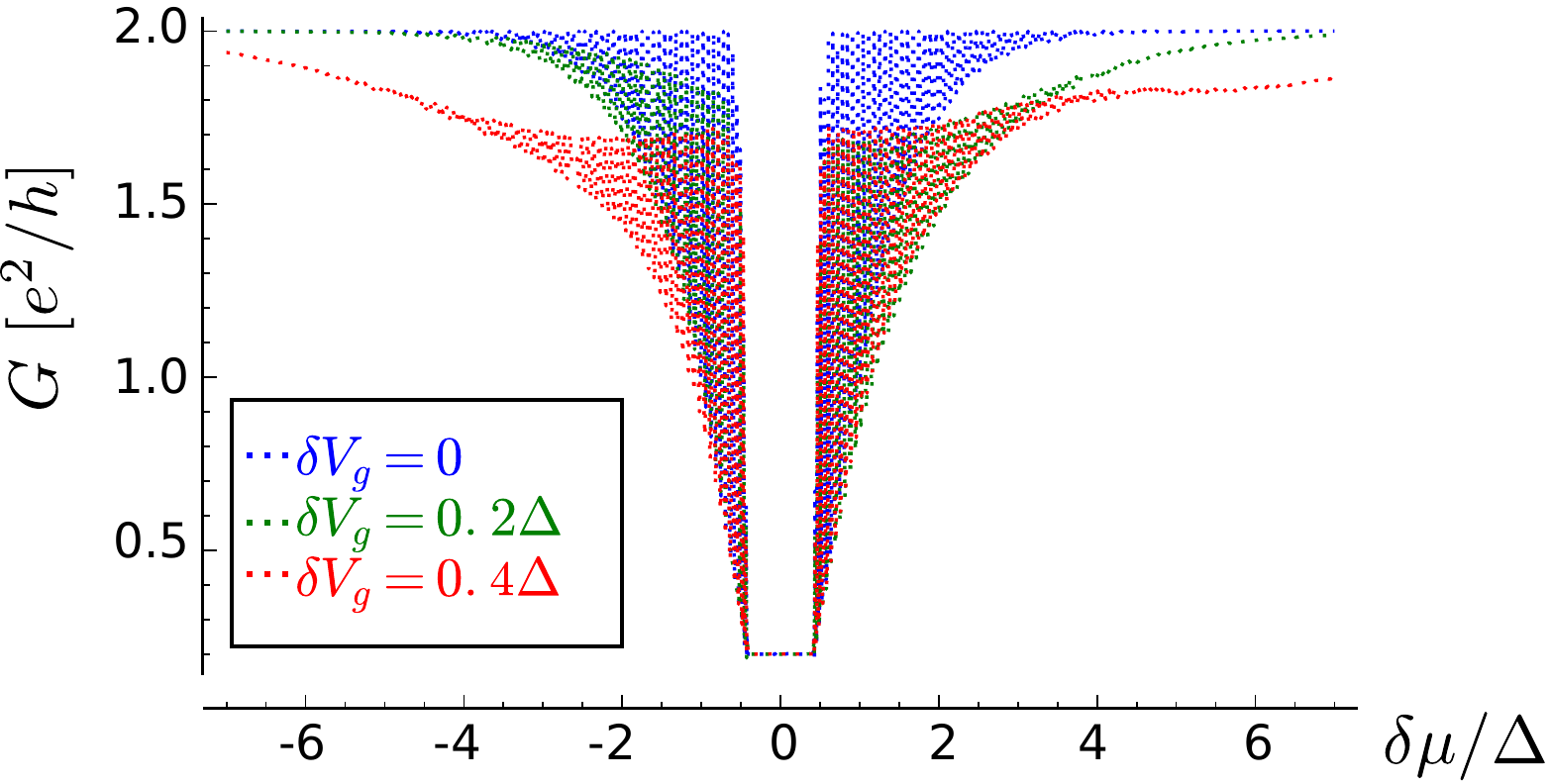}
	\caption{		
The same as Fig.~\ref{fig:G-vs-mu-adiabatic} but for an abrupt drop of the gate voltage at contacts, $l=l_V=0.1 \hbar s/\Delta$. If the shift of the chemical potential is less than $\mu^*\approx 0.3\Delta$, the fractional conductance is observed. For higher values of the shift, the conductance is still suppressed in comparison to the quantized conductance $2G_0 = 2e^2/h$. This suppression is caused by a mismatch of modes in the leads and in the wire, similarly to the one observed in Ref.~\onlinecite{RainisPRB2014} for non-interacting systems. The reflection at the contacts gives rise to well-pronounced Fabry-Perot oscillations with a period of order of $\hbar s/L$. 
}
\label{fig:G-vs-mu-abrupt}
\end{figure}

In addition, we also study a more realistic model of the contact taking into account a gate potential drop. We assume the gate potential has a smooth profile along the wire  (see Fig.~\ref{fig:Landau-Zener}) and is given by 
	\begin{align}
	V_g = V_g^{lead} - \frac{\delta V_g}{2}\left[\tanh \frac{x}{l_V} + \tanh \frac{L-x}{l_V}\right].
	\label{eqn:gate-voltage}
	\end{align}
	Here we disregard the voltage bias between the leads assuming that it is much smaller than the gap $\Delta$ and the gate voltage variation $\delta V_g$. The model is similar to the one used in Ref.~\onlinecite{RainisPRB2014}. The potential varies from the value $V_g^{lead}$ in the leads to the value $V_g^{lead} - \delta V_g$ in the wire and exhibits a linear behavior with the slope $\delta V_g / l_V$ around $x=0,L$. 
	
To take into account the position-dependent gate voltage $V_g(x)$ we replace the chemical potential $\mu$ in Eqs.~(\ref{eqn:refermion-away1})--(\ref{eqn:refermion-away2}) with $\mu(x) = V_g(x)$ such that the momentum shift $K$ is now given by the following expression:
\begin{align}
K(x) = \frac{6\delta \mu}{v_F} - \frac{3\delta V_g}{v_F}\left[\tanh \frac{x}{l_V} + \tanh \frac{L-x}{l_V}\right].
\end{align}
Next, we solve the system of differential equations [see Eqs.~(\ref{eqn:refermion-away1})--(\ref{eqn:refermion-away2})] numerically. The conductance in the limit of nearly adiabatic transition $l_V \gg \hbar s/ \Delta$ is shown in Fig.~\ref{fig:G-vs-mu-adiabatic}. If the gate potential variation $\delta V_g$ is smaller than $\mu^*$ (see Fig.~\ref{fig:Landau-Zener}a), the half-width of the dip in conductance remains the same, as well as the quantized conductance value, $G(|\delta \mu| \lesssim \mu^*) = e^2/5h$, is observed. However, if the total gate potential variation $\delta V_g$ becomes larger than  $\mu^*$, the dip in the measured conductance broadens and exceeds $2\mu^*$ (see Fig.~\ref{fig:Landau-Zener}b). This size of the dip can be estimated as $\mu^*+\delta V_g$. Thus, in order to measure the size of the gap opened by interactions, it is important to work in the regime of small gate potential drop. In addition, as  $\delta V_g$ gets larger, the Fabry-Perot oscillations become more pronounced.  In the opposite limit of abrupt transition $l_V \ll \hbar s/\Delta$, the gate potential drop $\delta V_g$ does not affect the gap~(see Fig.~\ref{fig:G-vs-mu-abrupt}), however, it leads to a decrease of conductance outside the gap, since the modes inside the wire and outside the wire do not match well in the presence of an effective barrier created by a gate voltage (see also discussion in Ref.~\onlinecite{RainisPRB2014}).

	\section{Semiclassical approximation \label{sec:semiclassical}}
	
In this section we treat the sine-Gordon action $S_E$ [see Eqs.~(\ref{eqn:action-gapped})--(\ref{eqn:action-coupling})] semiclassically~\cite{Rajaraman1982,Sakita1985}. We expand it around a static classical field configuration using the procedure developed in Refs.~\onlinecite{Aseev2017} and \onlinecite{BraunPRB1996}. This approach can be justified when fluctuations of the phase $\phi_1$ are small, {\it i.e.}, in case of strong electron-electron repulsion. Since the RG study shows that for $K_c < 3/\gamma_c^2$ the scaling dimension $D$ of the sine-Gordon term flows to zero~(see~Appendix~\ref{app:RG}), we assume that this is the case.
	
	The Euler-Lagrange equations for the action $S_E$ defined in Eqs.~(\ref{eqn:action-gapped})--(\ref{eqn:action-coupling}) read as
	\begin{align}	
	&\partial_\tau^2 \phi_1 + \partial_x\left( v_{1}^2\partial_x \phi_1\right) - \partial_x\left(v_{12}^2 \partial_x \phi_{2}\right)\nonumber\\
	 & \hspace{90pt}= \sqrt{2}\lambda \omega_0^2  \sin \left(\sqrt{2}\lambda \phi_1\right),\\
	&\partial_\tau^2 \phi_2 -\partial_x\left( v_{12}^2\partial_x \phi_1\right) + \partial_x\left( v_{2}^2 \partial_x \phi_{2}  \right) = 0. 
	\label{eqn:Euler-Lagrange-2}
	\end{align}
A static solution can be found from Eq.~(\ref{eqn:Euler-Lagrange-2}),
	\begin{align}
	\partial_x \phi_2 = \frac{v_{12}^2}{v_{2}^2}\partial_x \phi_1.
	\end{align}
The resulting equation for $\phi_1$ resembles the sine-Gordon equation~\cite{Sakita1985}, 
	\begin{align}
	\partial_x \left(\frac{\lambda}{K_c^2 \gamma_c^2 + 1}\partial_x \phi_1 \right) = \sqrt{2} \frac{\omega_0^2(x)}{v_F^2} \sin\left(\sqrt{2}\lambda \phi_1 \right).
	\end{align}
Next, similarly to the conventional sine-Gordon model~\cite{Sakita1985,BraunPRB1996}, we assume that the Hilbert space of the model consists of the vacuum sector (with vacuum state and its excitations) as well as of the sectors with different number of solitons (kinks) and their scattering states. The model defined by Eqs.~(\ref{eqn:action-gapped})--(\ref{eqn:action-coupling}) allows for a high number of solitons being activated at non-zero temperatures. We assume that at temperatures of the order of the gap $\Delta$ the soliton gas is still dilute such that the interactions between solitons can be disregarded.

The  classical static vacuum solution is trivial, $\phi_1^{0} = \phi_2^{0} = 0$. If the contact to the leads is adiabatic such that $K_c(x)$ varies slowly with $x$, the static solution for $\phi_1$ corresponding to an (anti-) kink localized at position $\xi$ can be written as 
	\begin{align}
	\check{\phi}^{K(A)}(x,\xi) = \begin{pmatrix} 1\\v_{12}^2/v_{2}^2 \end{pmatrix} \frac{2\sqrt{2}}{\lambda}\arctan \exp \left(\pm\int\limits_\xi^{x} \frac{x'}{\delta_0(x')}dx'  \right),
	\label{eqn:static-kink}
	\end{align}
	with $\delta_0^{-2}(x) = 2\omega_0^2(x)\left[1+K_c^2(x)\gamma_c^2\right]/v_F^2$. Here and below we use a short-hand notation $\check{\phi} = (\phi_1, \phi_2)^T.$ The upper sign in Eq.~(\ref{eqn:static-kink}) corresponds to a  kink solution, while the lower sign corresponds to an anti-kink  solution. 	
	
	We describe a configuration of the $N$-soliton gas by collective coordinates $\bm{\xi} = \left\{ \xi_m \right\}_{m=1}^N$ and labels $\bm{l} = \left\{ l_m \right\}_{m=1}^N$, where $l_m = K,A$ depending on whether the $m$-th soliton is a kink (K) or an anti-kink (A). The asymptotic form of the classical solution is given by
	\begin{align}
	\check{\phi}_{\bm{l},\bm{\xi}}(x) = \sum\limits_{m=1}^N \check{\phi}^{l_m}(x,\xi_{m}),
	\label{eqn:classical-gas}
	\end{align}
where $\check{\phi}^{K(A)}(x,\xi_m)$ is a classical solution for an (anti-) kink located at position $\xi_m$ and is given by~Eq.~(\ref{eqn:static-kink}). In the vicinity of a classical solution $\check{\phi}_{\bm{l},\bm{\xi}}$, we expand the fields as a sum of classical solutions and fluctuations $\delta \check{\varphi}(x,\tau)$  around them,
	\begin{align}
	\check{\phi}(x,\tau) = \check{\phi}_{\bm{l},\bm{\xi}}(x) +  \delta \check{\varphi}(x,\tau),
	\label{eqn:expansion}
	\end{align}
	and treat the center of the kink as dynamical variable $\bm{\xi}(\tau)$.
	
	 The correlator of the bosonic fields can be expressed by using functional integration over the fluctuations in the vicinity of the classical static solutions,
	 \begin{align}
	& \left\langle \phi_i(x,\tau) \phi_j(x,0) \right\rangle = \frac{1}{\mathcal{Z}}\sum\limits_{N=0}^{\infty}
	 e^{-NE_0/T} \left\langle \phi_i(x,\tau) \phi_j(x,0) \right\rangle_N,\\
 	 &\left\langle \phi_i(x,\tau) \phi_j(x,0) \right\rangle_N =\sum_{\bm{l}} \int\mathcal{D}\delta\check{\varphi}\;  \phi_i(x,\tau)\phi_j(x,0) \nonumber \\ 
 	 &\hspace{120pt} \times e^{-S_E[\check{\phi}_{\bm{l},\bm{\xi}}+\delta \check\varphi]} ,
 	 \label{eqn:functional-formulation}
	 \end{align}	
	where $E_0$ is the bare rest energy of one soliton, and the partition function $\mathcal{Z}$ is given by
	\begin{align}
	\mathcal{Z} = \sum\limits_{N=0}^\infty\sum_{\bm{l}} \int \mathcal{D}\delta\check{\varphi}\; e^{-S_E[\check{\phi}_{\bm{l},\bm{\xi}}+\delta \check\varphi]-NE_0/T} .
	\label{eqn:partition}
	\end{align}

We note that the representation defined in Eq.~(\ref{eqn:expansion}) is redundant: shifts of both the collective coordinate $\xi_m$ and the Goldstone zero-mode $\delta \check{\varphi} \propto \partial_{\xi_m} \check{\phi}^{l_m}(x, \xi_m)$ describe the same translation of the $m$-th soliton. In order to avoid double counting we have to perform the integration only over the fluctuations orthogonal to the zero-modes, {\it i.e.}, $\int dx\ \delta \check{\varphi}(x,\tau)\partial_{\xi_m} \check{\phi}^{l_m}(x, \xi_m)= 0$. This can be done by the Faddeev--Popov technique~\cite{GervaisSakitaPRD1975,Sakita1985, FaddeevPRB1967}. The integrals over the fluctuations around the static kink solutions in~Eq.~(\ref{eqn:functional-formulation}) should be understood as
	\begin{multline}
	\int \mathcal{D}\delta\check{\varphi}\; \mathcal{O} e^{-S_E[\check{\phi}_{\bm{l},\bm{\xi}}+\delta \check\varphi]} \to\\ \int \mathcal{D}\delta \check{\varphi} \prod\limits_{m=1}^N\mathcal{D}\xi_m\;\mathcal{O} \delta\left(Q_m[\xi_m]\right)
	\det \left(\frac{\delta Q_m}{\delta \xi_m}\right)\\
	\times e^{-S_E[\check{\phi}_{\bm{l},\bm{\xi}}+\delta \check\varphi]},
	\label{eqn:Green-gas} 
	\end{multline}
with the Faddeev-Popov functional defined as $Q_m[\xi_m] = \int dx\; \check{\phi}(x,\tau)\partial_x \check{\phi}^{l_m}(x,\xi_m)$.

	Using the expansion given by~Eq.~(\ref{eqn:expansion}), we can represent the correlator $\left\langle \phi_i(x,\tau) \phi_j(x,0) \right\rangle_N$ defined in~Eq.~(\ref{eqn:functional-formulation}) as sum of a contribution from solitons $\left\langle \phi_i(x,\tau) \phi_j(x,0) \right\rangle_s$,  and a contribution from ``mesons'', {\it i.e.}, ``background'' fluctuations, which we denote by $\left\langle \phi_i(x,\tau) \phi_j(x,0) \right\rangle_b$,  
	\begin{align}
	&\left\langle \phi_i(x,\tau) \phi_j(x,0) \right\rangle_N = \left\langle \phi_i(x,\tau) \phi_j(x,0) \right\rangle_{N,s} \label{eqn:correlator-expansion} \\
	&\hspace{110pt}+ \left\langle \phi_i(x,\tau) \phi_j(x,0) \right\rangle_{N,b}, \nonumber\\
	&\left\langle \phi_i(x,\tau) \phi_j(x,0) \right\rangle_{N,s} = \sum\limits_{m',m}
	\left\langle \phi^{l_{m'}}_i(x,\xi_{m'}) \phi^{l_{m}}_j(x,\xi_{m'}) \right\rangle_N\label{eqn:soliton-contribution} ,\\
	&\left\langle \phi_i(x,\tau) \phi_j(x,0) \right\rangle_{N,b} =
	\left\langle \delta\phi_{i}(x,\tau) \delta \phi_{j}(x,0) \right\rangle_N\, .
	\label{eqn:background-fluctuations}
	\end{align}
We note that the cross-terms being odd in $\delta \varphi$ vanish while averaging over fast meson modes.

The charge current should not depend on the coordinate. Moreover, it is convenient to take $x$ in Eqs.~(\ref{eqn:correlator-expansion})--(\ref{eqn:background-fluctuations}) inside the left lead, {\it i.e.} $x<0$. In this case, $K_c(x) = 1$ and $v_{12}(x) = 0$, so the classical solution of the Euler-Lagrange equations for $\phi_2$ is trivial, $\phi_2=0$. Thus, the only non-zero correlator in Eq.~(\ref{eqn:soliton-contribution}) is the one with $i=j=1$.  As a result, according to~Eq.~(\ref{eqn:Kubo}), the $N$-soliton state contributes to the conductance as 
	\begin{align}
	&G_N =\lim\limits_{\bar{\omega}\to0}\bar{\omega}\left\{ \frac{2e^2 \gamma_c^2}{\pi^2 (\gamma_c^2+1)} \langle \phi_1(x,\bar{\omega}) \phi_1(x,-\bar{\omega}) \rangle_{N,s} \right. 	\label{eqn:Kubo-extended}\\ 
	&\hspace{20pt}+
	\frac{2e^2 }{\pi^2 (\gamma^2_c+1)} \left[\gamma_c^2 \langle \delta \phi_1(x,\bar{\omega}) \delta \phi_1(x,-\bar{\omega}) \rangle_N\right. \nonumber\\
	&\left. + \gamma_c \langle \delta \phi_1(x,\bar{\omega}) \delta \phi_2(x,-\bar{\omega}) \rangle_N + \gamma_c\langle \delta \phi_2(x,\bar{\omega}) \delta \phi_1(x,-\bar{\omega})\rangle_N\right.\nonumber\\
	&\hspace{20pt} \left. \left. + \langle \delta \phi_2(x,\bar{\omega}) \delta \phi_2(x,-\bar{\omega}) \rangle_N  \right]\right\}\, .\nonumber
	\end{align}
	
In the following we show that the contribution from correlators for background fluctuations $\langle \delta \phi_i \delta \phi_j \rangle$ yields a temperature independent fractional conductance $2e^2/(\gamma_c^2+1)h$ [see Secs. \ref{subsec:vacuum}, \ref{subsec:background}]. In contrast to that, as shown in Sec.~\ref{subsec:one-kink}, the conductance acquires a temperature dependence already in the one-kink approximation $N=1$. In  Sec.~\ref{subsec:gas}, we generalize this result in the dilute soliton gas limit.

	\subsection{Vacuum sector \label{subsec:vacuum}}
First, we calculate the correlators $\left\langle \phi_i(x,\tau) \phi_j(x,0) \right\rangle$ in Eq.~(\ref{eqn:correlator-expansion}) for the vacuum sector, {\it i.e.} for $N=0$. We insert an auxiliary point source term $\left(\mathscr{J}_1 \delta \varphi_1 + \mathscr{J}_2 \delta \varphi_2\right)  \delta(x-x')$ into the action $S_E$, then the solution of the Euler-Lagrange equations in real time can be represented as $\delta \varphi_\alpha(x) = -i\mathcal{G}^R_{\alpha\beta}(x,x') \mathscr{J}_\beta $, where $\mathcal{G}_{\alpha \beta}^R(x,x')$ is a retarded Green function. In order to calculate the conductance in the left lead we assume that the sources $\mathcal{J}_{1,2}$ are also located in the left lead, $x'<0$.  The Euler-Lagrange equations for small fluctuations $\delta \check{\varphi}$ are given by 
\begin{align}
&\omega^2 \delta \varphi_1 + \partial_x \left( v_{1}^2(x) \partial_x \delta\varphi_1 + v_{12}^2(x)\partial_x \delta\varphi_2  \right)-W^2(x) \delta\varphi_1 \nonumber \\
& \hspace{90pt}= \pi v_F \mathscr{J}_1 \delta(x-x'),
	\label{eqn:Euler-Lagrange-fluc1}\\
&	\omega^2 \delta \varphi_2 + \partial_x \left( v_{2}^2 \partial_x \delta\varphi_2 + v_{12}^2\partial_x \delta\varphi_1  \right)	  = \pi v_F \mathscr{J}_2 \delta(x-x').
	\label{eqn:Euler-Lagrange-fluc2} 
\end{align}
The potential $W(x) = 2\lambda^2\omega_0^2(x)e^{-\lambda^2\langle \phi_1^2(x) \rangle}$ obtained by extending the action around a static vacuum solution in the self-consistent harmonic-approximation~\cite{GiamarchiBook}. In following consideration, we do not assume any specific shape of the potential $W(x)$ except that it is localized inside the wire and vanishes in the leads $W(x>L) = W(x<0) = 0$.  The velocities $v_{1}(x)$, $v_{2}(x)$ coincide with the Fermi velocity if $x$ is outside the wire, and the coupling $v_{12}$ vanishes in the leads, $v_{12}(x<0) = v_{12}(x>L) = 0$. 
	
To begin,	we study the scattering problem defined by  Eqs.~(\ref{eqn:Euler-Lagrange-fluc1})--(\ref{eqn:Euler-Lagrange-fluc2}) with zero sources $\mathcal{J}_1 = \mathcal{J}_2=0$. There are two solutions corresponding to the wave incident from the left lead, which have the following asymptotic form,
	\begin{align}
	&\begin{pmatrix}
	\delta \varphi_1\\
	\delta \varphi_2
	\end{pmatrix}_1 = 
\begin{cases}
\begin{split}
	e^{i\frac{\omega x}{v_F}} \begin{pmatrix}
	1\\0
	\end{pmatrix} + r_{11}e^{-i\frac{\omega x}{v_F}} \begin{pmatrix}
	1\\0
	\end{pmatrix} \\+
		r_{21}e^{-i\frac{\omega x}{v_F}} \begin{pmatrix}
	0\\1
	\end{pmatrix}&,\quad x<0\end{split} \\
	t_{11}e^{i\frac{\omega x}{v_F}} \begin{pmatrix}
	1\\0
	\end{pmatrix} +
	t_{21}e^{i\frac{\omega x}{v_F}} \begin{pmatrix}
	0\\1
	\end{pmatrix}&,\quad x>L,
\end{cases}
	\label{eqn:uniform-sol1}\\
&	\begin{pmatrix}
	\delta \varphi_1\\
	\delta \varphi_2
	\end{pmatrix}_2 = 
	\begin{cases}
	\begin{split}
	e^{i\frac{\omega x}{v_F}} \begin{pmatrix}
	0\\1
	\end{pmatrix} + r_{22}e^{-i\frac{\omega x}{v_F}} \begin{pmatrix}
	0\\1
	\end{pmatrix} \\+
	r_{12}e^{-i\frac{\omega x}{v_F}} \begin{pmatrix}
	1\\0
	\end{pmatrix}&,\quad x<0\end{split}\\
	t_{12}e^{i\frac{\omega x}{v_F}} \begin{pmatrix}
	1\\0
	\end{pmatrix} +
	t_{22}e^{i\frac{\omega x}{v_F}} \begin{pmatrix}
	0\\1
	\end{pmatrix}&,\quad x>L,
	\end{cases}	
		\label{eqn:uniform-sol2}
	\end{align} 
	where $t_{ij}(\omega)$ are transmission amplitudes for scattering from the $j$-th mode in the left lead to the $i$-th mode in the right lead, and $r_{ij}(\omega)$ are reflection amplitudes for scattering from $j$-th mode in the left lead to the $i$-th mode in the same lead. Thus, the solution of~Eqs.~(\ref{eqn:Euler-Lagrange-fluc1})--(\ref{eqn:Euler-Lagrange-fluc2}) can be reduced to a scattering problem. In the limit $\omega \to 0$, the potential barrier $W(x)$ becomes impenetrable for the first mode $\delta \varphi_1$, and $r_{11}(\omega) = -1 + O(\omega)$, $r_{12}(\omega) = r_{21}(\omega) = r_{22}(\omega)= O(\omega)$ (the derivation of these asymptotics is similar to the one given in Ref.~\onlinecite{LandauLifshitz3}).
	
	Now we proceed with the solution of Eqs.~(\ref{eqn:Euler-Lagrange-fluc1})--(\ref{eqn:Euler-Lagrange-fluc2}) with non-zero sources. The solution at $x<x'<0$ is a wave propagating from the source at $x'$ to the left and is of the form
	\begin{align}
	\begin{pmatrix}
	A_1\\A_2
	\end{pmatrix} e^{-i\frac{\omega (x-x')}{v_F}}.	
	\end{align}
The solution at $x'<x<0$ is a linear combination of the solutions given in Eqs.~(\ref{eqn:uniform-sol1})--(\ref{eqn:uniform-sol2}),
	\begin{align}
	B_1\begin{pmatrix} \delta \varphi_1\\ \delta \varphi_2
	\end{pmatrix}_1 + B_2 \begin{pmatrix} \delta \varphi_1\\ \delta \varphi_2
	\end{pmatrix}_2.
	\end{align}
	The matching conditions at $x=x'$ read
	\begin{align}
	&A_1 = B_1\left( e^{\frac{i\omega x'}{v_F}} + r_{11}e^{\frac{-i\omega x'}{v_F}}  \right) + B_2r_{12}e^{-\frac{i\omega x'}{v_F}},\\
	&A_2 = B_2\left( e^{\frac{i\omega x'}{v_F}} + r_{22}e^{\frac{-i\omega x'}{v_F}}  \right)+ B_1r_{21}e^{-\frac{i\omega x'}{v_F}},\\
	&B_1\left( e^{\frac{i\omega x'}{v_F}} - r_{11}e^{\frac{-i\omega x'}{v_F}}  \right) - B_2r_{12}e^{-\frac{i\omega x'}{v_F}} + A_1 = \frac{\pi \mathcal{J}_1}{i \omega},\\
	&B_2\left( e^{\frac{i\omega x'}{v_F}} - r_{22}e^{\frac{-i\omega x'}{v_F}}  \right)- B_1r_{12}e^{-\frac{i\omega x'}{v_F}} + A_2 = \frac{\pi \mathcal{J}_2}{i \omega}.
\end{align}
Solving this system of linear equations, we obtain the following response $\delta \varphi_{i}$ to the source term, calculated at $x=x'$:
\begin{align}
\begin{pmatrix}
\delta\varphi_1\\
\delta \varphi_2
\end{pmatrix}
=
\frac{\pi}{2i \omega} 
\begin{pmatrix}
\mathcal{J}_1\left(1 + r_{11}e^{-2i\omega\frac{x'}{v_F}} \right) + \mathcal{J}_2r_{12}e^{-2i\omega\frac{x'}{v_F}}\\
\mathcal{J}_1r_{21}e^{-2i\omega\frac{x'}{v_F}} +
\mathcal{J}_2\left(1 + r_{22}e^{-2i\omega\frac{x'}{v_F}} \right)
\end{pmatrix}.
\end{align}
Thus, the retarded Green functions are of the following form:
\begin{align}
\mathcal{G}^R_{11}(x',x',\omega) &=\; \frac{\pi}{2(\omega+i0)}\left(1 + r_{11}e^{-2i\omega\frac{x'}{v_F}} \right)
= O(1),\\
\mathcal{G}^R_{12}(x',x',\omega) &=\; \frac{\pi}{2(\omega+i0)}r_{12}
= O(1),\\
\mathcal{G}^R_{21}(x',x',\omega) &=\; \frac{\pi}{2(\omega+i0)}r_{21}
= O(1),\\
\mathcal{G}^R_{22}(x',x',\omega) &=\; \frac{\pi}{2(\omega+i0)}\left(1 + r_{22}e^{-2i\omega\frac{x'}{v_F}} \right) \nonumber
\\&=\; \frac{\pi}{2(\omega+i0)} + O(1).
\end{align}
Now we perform an analytical continuation of the retarded Green functions to obtain the correlators from Eq.~(\ref{eqn:Kubo-extended}) in Matsubara representation.
According to Eq.~(\ref{eqn:Kubo-extended}),  the only non-zero contribution to the conductance is obtained from $\mathcal{G}^R_{22}$, yielding 
\begin{align}
G_{vac} = \frac{1}{\mathcal{Z}}\frac{2G_0}{\gamma_c^2+1}.
\label{eqn:conductance-vac}
\end{align}
 In the limit of low temperatures $T\to 0$, $\mathcal{Z} = 1$, and we obtain the known result for the low-temperature fractional conductance\cite{OregPRB2014, MengPRB2014}, which is also in agreement with the results obtained in Sec.~\ref{sec:refermionization-T0} [see Eq.~(\ref{eqn:fractional})],
\begin{align}
G(T=0) = \frac{2G_0}{\gamma_c^2+1}.
\end{align}
	
	\subsection{Contribution to the conductance from background fluctuations for $N\ge 1$ \label{subsec:background}}
	The effective action for fluctuations $\delta \varphi$ in the presence of $N$ solitons reads
	\begin{multline}
	S_{\delta \varphi} = \frac{1}{2\pi v_F} \int dx d\tau\ \delta \varphi_1 \left\{-\partial_\tau^2 - \partial_x v_{1}^2 \partial_x +W_N(x) \right\} \delta \varphi_1 \\+ S_2[\delta \varphi_2] + S_{12}[\delta \varphi_1, \delta\varphi_2],
	\end{multline}
	where $W_N(x)$ is the effective potential created by $N$ solitons,
	\begin{align}
	W_N(x) = \frac{1}{2}\left.\left(\frac{\delta^2}{\delta \phi_1^2}\; \left[ \omega_0^2\cos\left(\sqrt{2}\lambda \phi_1  \right)\right]\right)\right|_{\check{\phi} =  \check{\phi}_{\bm{l},\bm{\xi}}}. 
	\end{align}
	We recall that the analysis in the previous section does not rely on a particular form of the potential $W(x)$. Thus, the correlator for background fluctuations is given by
	\begin{align}
	\langle \delta\varphi_i(x,\bar{\omega}) \delta\varphi_j(x,0) \rangle_N = \frac{\pi}{2 \bar{\omega}}.
	\end{align}
Combining~Eq.~(\ref{eqn:Kubo-extended}) with~Eq.~(\ref{eqn:functional-formulation}), we obtain the contribution to the conductance  $G_{bg, N}$ due to background fluctuations in the presence of $N$ solitons,
	\begin{align}
	G_{bg, N} = \frac{1}{\mathcal{Z}}\frac{2G_0}{\gamma_c^2+1} e^{-NE_0/T}.
	\end{align}
This expression generalizes Eq.~(\ref{eqn:conductance-vac}) obtained for  vacuum fluctuations. Summing up all the contributions from $N=0$ up to $N\to\infty$, we arrive at 
$\sum_N G_{bg, N} = 2G_0/(\gamma_c^2 + 1)$. This contribution turns out to be temperature-independent and coincides with the fractional conductance $ G_\nu$.

	\subsection{One-kink approximation\label{subsec:one-kink}}
In this subsection we focus on the low-temperature regime, $T\ll \Delta$, and consider only states with not more than one soliton. Consenquently, we  omit terms with $N>1$ in Eq.~(\ref{eqn:partition}). For the sake of simplicity, in what follows we do not track how the bare rest energy $E_0$ is renormalized by fluctuations into a physical gap $\Delta$. We estimate the value of the gap from the first-order RG equations, see Eqs.~(\ref{eqn:1st-RG})--(\ref{eqn:1st-RG-c}).

At finite temperature, a kink with the renormalized rest energy $\Delta$  and the mass $M\sim \Delta/s^2$ can be activated. The kink propagates inside the wire carrying electric charge and interacting with the environment consisting of gapless and gapped modes of background fluctuations (see Appendix~\ref{app:eff-action} for details). The spectrum of the fluctuation modes is given by
\begin{align}
\left(\omega_q^{\pm}\right)^2 = \frac{\omega_0^2+2 q^2 v_{11}^2\pm\sqrt{\left[\omega_0^2 + \left(v_{1}^2-v_{2}^2 \right)q^2 \right]^2 + 4v_{12}^4q^4 }}{2}.
\end{align}
	The plus sign corresponds to the gapped mode $\omega_q^{+}\approx \sqrt{\omega_0^2 +q^2 v_{1}^2}$, while the minus sign corresponds to a gapless acoustic mode $\omega_q^- \approx v_{2}|q|$ for $qv_{2} \ll \omega_0$. The gapped mode leads to a renormalization of the kink rest energy\cite{GervaisSakitaPRD1975, Sakita1985}. 
A coupling to gapless modes causes an effective friction such that the kink dissipates energy by interacting with the gapless mesons. This mechanism resembles Caldeira-Leggett type dissipation~\cite{CaldeiraLeggett1983}  and damping of Bloch walls in quasi-1D ferromagnets caused by interaction with spin waves~\cite{BraunPRB1996}.
	
In order to calculate a contribution to the conductance due to the motion of kinks we integrate out the fluctuations $\delta \check{\varphi}$ and obtain an effective low-energy Euclidean action for the collective coordinate $\xi$ (see Appendix~\ref{app:eff-action} and Ref.~\onlinecite{Aseev2017} for details of the derivation),
	\begin{align} 
	S_{eff}[\xi] = T\sum\limits_{\bar \omega}\; \xi\left(\bar{\omega}\right)\left\{\frac{M \bar{\omega}^2 }{2} + \frac{M}{2}\eta |\bar{\omega}|
	\right\}\xi\left(-\bar{\omega}\right),
	\label{eqn:effective-action}
	\end{align}
where the summation over Matsubara frequencies $\bar{\omega}$ is performed. The first term describes the free motion of the kink, while the second term corresponds to an 'Ohmic-like' friction ({\it i.e.}, linear in $|\bar{\omega}|$) caused by the interaction between the kink and the gapless fluctuation modes. The temperature-dependent friction coefficient is given by
	\begin{align}
	\eta &= \; 4\frac{T}{\Delta} \frac{s^2}{\delta_0}\frac{v_{12}^2}{v_{2}^3}.  
	\label{eqn:friction}
	\end{align}	
	
The Matsubara Green function for the collective coordinate $\mathcal{D}^M(\bar{\omega})$ for long wires in the limit $L \gg s/\Delta, s/T$ reads 
	\begin{align}
	\mathcal{D}^{M}(\bar{\omega}) &=\; \frac{1}{ M\left( \bar{\omega}^2 + \eta |\bar{\omega}|\right)}.
	\end{align}	
	The retarded Green $\mathcal{G}^R$ function for the bosonic field $\phi_i$ can be related to the Green functions $\mathcal{D}$ for collective coordinates as was shown in  Ref.~\onlinecite{Aseev2017}:	
	\begin{multline}
	\mathcal{G}^R(x,x,\tau) =  4\pi \Theta(\tau)\sqrt{\frac{M}{\beta}} \dfrac{\mathcal{D}^R(\tau)-\mathcal{D}^R(0)}{\sqrt{i(\mathcal{D}^K(0)- \mathcal{D}^K(\tau))}}\\\times \dfrac{\sqrt{2}}{\sqrt{\sqrt{1-\dfrac{\left(\mathcal{D}^R(\tau)-\mathcal{D}^R(0)\right)^2}{(\mathcal{D}^K(\tau)-\mathcal{D}^K(0))^2}}+1}},
	\label{eqn:retarded}
	\end{multline}
	where $\mathcal{D}^K = -i\left\langle\left\{ \xi(t),\xi(t') \right\}\right\rangle$ and $\mathcal{D}^R = -i\Theta(t-t')\langle[\xi(t),\xi(t')]\rangle$ are Keldysh and retarded Green functions for collective coordinates, respectively, and $\beta=1/T$. The retarded Green function $\mathcal{D}^R $ can be extracted from the Matsubara Green function $\mathcal{D}^{M}$ by analytic continuation,
	\begin{align}
	\mathcal{D}^{R}(\omega) &=\; \frac{1}{ M\left(\omega + i0 \right)\left(\omega + i\eta \right)},\\ 
	\mathcal{D}^{R}(\tau) &=\; \frac{1}{M} \frac{1- e^{-\eta \tau}}{\eta}\Theta(\tau).
	\label{eqn:D-retarded-time}
	\end{align} 
The Keldysh Green function can be obtained using the fluctuation-dissipation theorem, 
	\begin{align}
	&\mathcal{D}^K(\omega) = 2\mathrm{Im}\; \mathcal{D}^R(\omega)\coth \frac{\omega}{2T} \nonumber\\
	&\hspace{80pt}= -\frac{2i}{M \omega}\frac{\eta}{\eta^2+\omega^2}\coth \frac{\omega}{2T}, \\
	&\mathcal{D}^K(\tau)-\mathcal{D}^K(0) \approx \frac{2i}{\pi M} T\tau^2 \arctan \frac{1}{\eta \tau}.
	\end{align}
As a result, in the absence of the friction the Keldysh Green function reads
	\begin{align}
	\mathcal{D}^K(\tau)-\mathcal{D}^K(0) = \frac{i}{M}T\tau^2.
	\end{align}	

	The conductance is related to low-frequency current--current correlator by~Eq.~(\ref{eqn:Kubo-extended}), and, hence, the contribution to the conductance from one kink, $G_K$, can be extracted from the retarded Green function in time-representation, 
	\begin{align}
	G_{K} = \frac{2\gamma_c^2}{\gamma_c^2+1}e^{-\Delta/T}\lim\limits_{ \tau \to+\infty} \mathcal{G}^R(\tau).
	\label{eqn:G-time-limit}
	\end{align}
	
The activation law exponent arises due to the rest energy $\Delta$ of the kink. First, we focus on the important limiting case of an infinitely long wire in the absence of friction, $\eta = 0$, $L \gg s/\Delta, s/T$. If one disregards dissipation terms in Eq.~(\ref{eqn:D-retarded-time}), the Green functions for the collective coordinate grows infinite with time,		
	\begin{align}
	&	\mathcal{D}^{R}(\tau)\vert_{\eta \to 0} =\; \frac{\tau}{M}\Theta(\tau),\\		
	&		\mathcal{D}^K(\tau)-\mathcal{D}^K(0)\vert_{\eta \to 0} =\; \frac{i}{M}T\tau^2.
	\end{align}
	The kink and equal anti-kink contribution to the conductance, $G_K$ and $G_A$, respectively, are obtained straightforwardly from~Eq.~(\ref{eqn:G-time-limit}),	\begin{align}
	{G}_{K}={G}_{A} = \frac{2G_0\gamma^2_c}{\gamma^2_c+1} e^{-\Delta/T},
	\end{align}
	so that the total conductance at $T\ll \Delta$ is $G = G_\nu + 2G_K$.	Thus, comparing the result with the asymptotics of Eq.~(\ref{eqn:G-T-dependence}) at $T\ll \Delta$, we see that in the absence of friction the interactions only change  slightly the temperature-dependence of the gap $\Delta(T)$ [see Eq.~(\ref{eqn:1st-RG-b})].
	
The situation changes drastically if the dissipation is taken into account, $\eta > 0$, however, the wire is still assumed to be long enough. Now the retarded Green function for the collective coordinate $D^R$ is finite at infinite times, $\mathcal{D}^R(\tau\gg \eta^{-1}) = 1/(M\eta)$, but the Keldysh Green function (in the limit of a long wire $L\gg v_F/\eta$) is still infinite, $\mathcal{D}^K(\tau\gg \eta^{-1}) \propto \tau^2$. Therefore, Eq.~(\ref{eqn:G-time-limit}) yields zero conductance. This can be easily understood, since the Ohmic-like friction causes an internal resistivity, and we may expect that in long wires, $L> v_F/\eta$, the total conductance will drop to zero as the wire length $L$ is increased. 
	
	\begin{figure}
	\includegraphics[width=\linewidth]{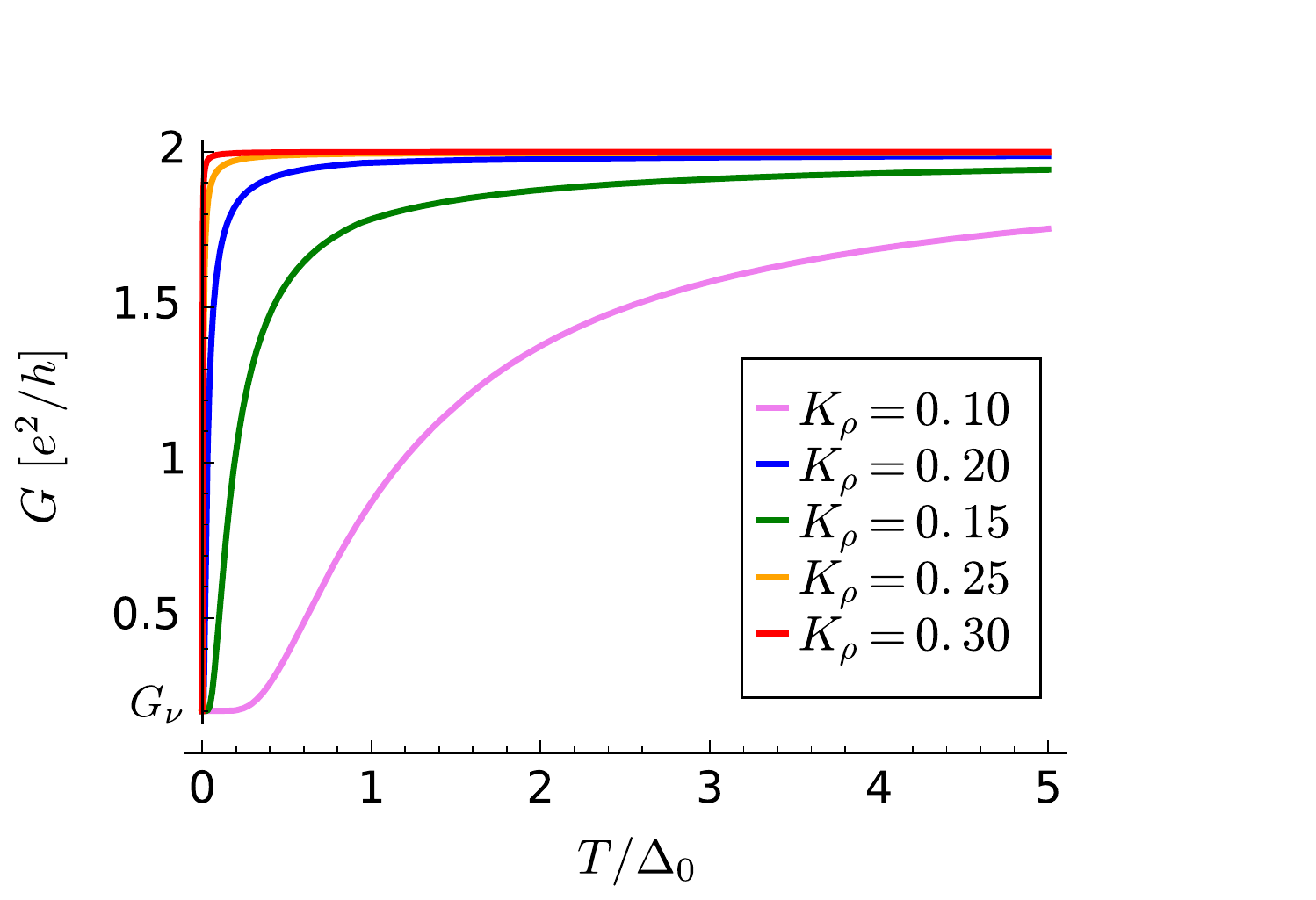}	
	\caption{Conductance of a long wire, $L\gg \hbar v_F/\Delta_0$, as function of temperature $T$ at filling factor $\nu=1/3$ assuming that friction experienced by solitons can be disregarded [see Eq.~(\ref{eqn:G-nofriction})]. The temperature is given in units of the bare (non-renormalized) gap $\Delta_0 \simeq BU_{2k_F}^2/v_F^2$, while the conductance is measured in units of  $G_0 = e^2/h$. At $T=0$ the conductance coincides with $G_\nu =e^2/5h$, while at high temperatures, $T\gg \Delta$, the conductance reaches its full value $2e^2/h$. The activation curve becomes steeper as the interaction parameter $K_c$ approaches  the critical value $K_c = 1/3$. The activation temperature is determined by the renormalized value of the gap $\Delta$ and is governed by~Eq.~(\ref{eqn:1st-RG-b}). The length of the wire was taken much longer than the correlation length $\hbar v_F/\Delta$.  }
	\label{fig:conductance_nofriction}
\end{figure}	
\begin{figure*}[!t]
	\includegraphics[width=\linewidth]{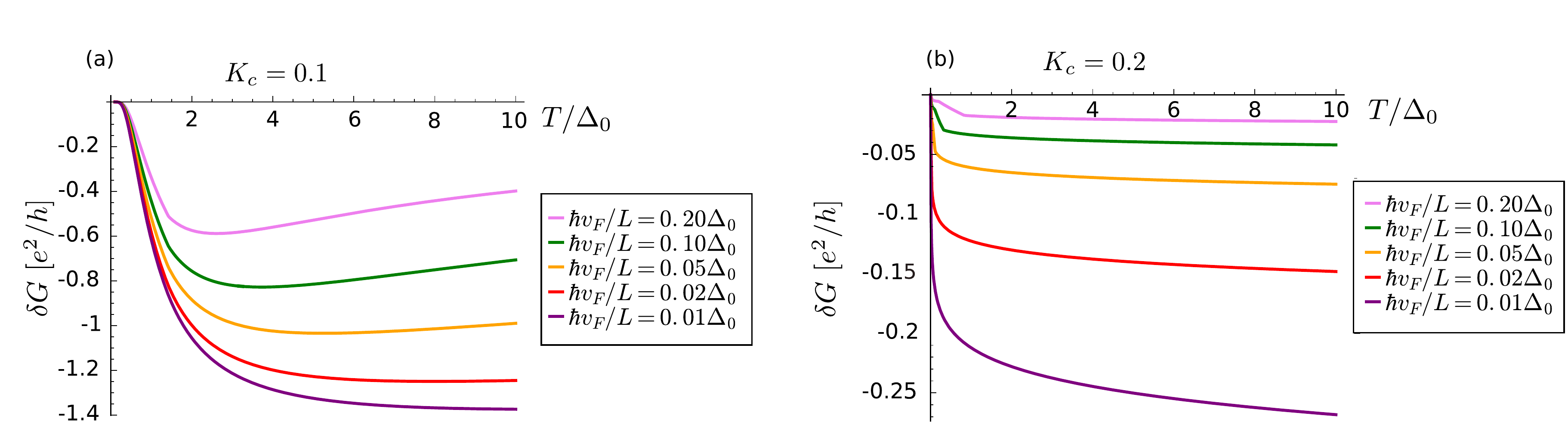}	
	\caption{Temperature dependence of the difference $\delta G = G - G_{\eta=0}$ between the total conductance $G$ at the filling factor $\nu=1/3$ with friction taken into account [see Eq.~(\ref{eqn:total-conductance})] and the conductance $G_{\eta=0}$ calculated disregarding effects of friction [see Eq.~(\ref{eqn:G-nofriction})] for two values of Luttinger liquid parameter: (a) $K=0.1$ and (b) $K=0.2$. Temperature is measured in units of the bare  gap $\Delta_0 \simeq BU_{2k_F}^2/v_F^2$, while the conductance is measured in units of $G_0 = e^2/h$. The temperature dependence of the gap is determined using~Eq.~(\ref{eqn:1st-RG-b}). The friction coefficient $\eta$ is estimated using~Eq.~(\ref{eqn:friction}).
At high temperatures $T>\Delta$, the conductance is suppressed by the friction. The correction to the conductance due to friction becomes more significant in case of strong interactions  [see panel (a)].  The effect of friction may be neglected for weak interaction, if the length of the wire $L$ is not much longer than the correlation length $\hbar v_F/\Delta$  [see panel (b)]. In contrast to that, if the wire is longer, the friction plays an important role also in this case.
			}
	\label{fig:conductance_frictionK0102}
\end{figure*}	

The crossover between these regimes can be roughly described by taking the limit at finite $\tau\to\tau_\infty$ instead of $\tau\to \infty$ in~Eq.~(\ref{eqn:G-time-limit}),
	\begin{align}
	G_{K} = \frac{2G_0\gamma^2_c}{\gamma^2_c+1} \sqrt{T M} \frac{\dfrac{1-e^{-\eta \tau_\infty}}{M \eta} }{\sqrt{\frac{2}{\pi M} T \tau_\infty^2 \arctan \frac{1}{\eta \tau_\infty}}} e^{-\Delta/T}.
	\label{eqn:one-soliton-result}
	\end{align} 
	
The results for finite but large wire length $L$ can be easily estimated. Since the collective coordinate is bounded inside the wire $0<\xi<L$, the Keldysh and retarded Green functions must be bounded as well, $|\mathcal{D}^K| < L^2$, $|\mathcal{D}^R| < L^2$.  Therefore, we assume that the Green functions grow until they reach their asymptotic value of order of $L^2$. This gives a cut-off parameter at large times $\tau_\infty = \min\{\tau_R, \tau_K\}$ with
	\begin{align}
	\tau_R =  \dfrac{\Delta L^2}{\hbar s^2},\,\,\,\, \tau_K = \dfrac{L}{s}\sqrt{\dfrac{\Delta}{T}}.
	\end{align}
First, we consider the limit of long cut-off time $\tau_\infty$ and of strong friction $\eta$ such that $\eta \tau_\infty \gg 1$. One can estimate the corresponding temperatures at which this regime occurs as $T\gg \hbar v_2^6/(s v_{12}^4L), \hbar^2 s v_2^3/(\Delta v_{12}^2 L^2)$. In this case the conductance is suppressed by friction,
	\begin{align}
	G_{K} = \frac{2G_0\gamma^2_c}{\gamma^2_c+1} \sqrt{\frac{T}{\Delta}}\frac{s}{\eta(T) L}e^{-\Delta/T},\quad \eta \tau_\infty \gg 1.
	\label{eqn:GK-friction}
	\end{align}
	In the opposite limit $\eta \tau_\infty \ll 1$, the friction becomes insignificant, and the conductance is the same as in the non-interacting case.

\subsection{Dilute soliton gas approximation \label{subsec:gas}}

Now we return to the general case of a soliton gas consisting of $N$ solitons (kinks and antikinks) with a classical configuration described by~Eq.~(\ref{eqn:classical-gas}).
Assuming that the soliton gas at temperatures $T\gtrsim \Delta$ is dilute, we disregard interactions between solitons. In this case, the effective action for the $N$-soliton gas can be written as
\begin{align}
S_E[\xi_1,\dots,\xi_N] = \sum\limits_{i=1}^N S_{eff}[\xi_i],
\end{align}
where $S_{eff}$ is given by~Eq.~(\ref{eqn:effective-action}). The integration over $\xi_k$ in~Eq.~(\ref{eqn:Green-gas}) can be reduced to a one-kink retarded Green function $\mathcal{G}^{R}_{N=1}$ for the action $S_{eff}[\xi]$  calculated in the previous section, see~Eq.~(\ref{eqn:retarded}). As a result, the summation over $N$ can be easily performed, 
\begin{align}
&\mathcal{Z} = \frac{1}{1+e^{-\Delta/T}}\, ,\\
&\mathcal{G}^R(t,t') = -2\frac{d \ln \mathcal{Z}}{d\left( \Delta/T\right)}\mathcal{G}^{R}_{N=1} = \frac{2e^{-\Delta/T}}{1+ e^{- \Delta/T}}\mathcal{G}^{R}_{N=1}.
\end{align}
In comparison to the one-kink approximation, the Green functions and, correspondingly, the conductance acquire an extra activation factor $e^{-\Delta/T}/\left( 1 + e^{-\Delta/T} \right)$ instead of $e^{-\Delta/T}$ in the previous section. In the limiting case when the decay length of solitons is less than the length of the wire, $\eta \tau_\infty \gg 1$,  the total conductance is given by 
\begin{align}
G= \frac{2G_0}{\gamma^2_c+1} \left(1 + \frac{2 e^{-\Delta/T}}{1+e^{-\Delta/T}}\sqrt{\frac{T}{\Delta}}\frac{s}{\eta L}\right)
\label{eqn:total-conductance-low}.
\end{align}
 The first term  stems from the background fluctuations, and does not depend on temperature. The second temperature-dependent term differs from~Eq.~(\ref{eqn:GK-friction}) by a temperature-activation factor which now takes into account summation over soliton gas configurations with different number  of solitons $N$.

In the opposite regime, when the solitons can almost freely propagate through the wire,  $\eta \tau_\infty \ll 1$, the friction becomes insignificant. As a result, the total conductance is given by the simple expression
\begin{align}
G = \frac{2G_0}{\gamma^2_c+1}\left(1 + \frac{2\gamma^2_c e^{-\Delta/T}}{1+e^{-\Delta/T}} \right).
\label{eqn:G-nofriction}
\end{align}
At zero temperatures, $T=0$, the result agrees with the fractional conductance $G_{\nu}= 2G_0/(\gamma^2_c+1)$ found before~\cite{OregPRB2014}. 

Finally, the expression for the conductance at the crossover between these regimes, which was given by~Eq.~(\ref{eqn:one-soliton-result}) in the one-soliton approximation, is now replaced by 
\begin{align}
G = 
G_{\nu}
\left(1+ \frac{2\sqrt{T M}e^{-\Delta/T}}{1+e^{-\Delta/T}}  
\frac{\dfrac{1-e^{-\eta \tau_\infty}}{M \eta} }{\sqrt{\frac{2}{\pi M} T \tau_\infty^2 \arctan \frac{1}{\eta \tau_\infty}}}\right).
\label{eqn:total-conductance}
\end{align}
The resulting conductance for $\eta = 0$ ({\it i.e.}, neglecting friction) and for different values of interaction parameters is shown in Fig.~\ref{fig:conductance_nofriction}. At zero temperature, the conductance is given by the   fractional quantum value $G_\nu$. At high temperatures, the conductance reaches its full value $2e^2/h$.  The interaction strength affects the temperature dependence of the renormalized gap [see Eq.~(\ref{eqn:1st-RG})] such that the stronger interaction is, the larger the gap, and, hence, the activation temperature is. 

The correction to the conductance due to the friction is shown in Fig.~\ref{fig:conductance_frictionK0102}.  The resistance due to friction vanishes at $T=0$. At higher temperatures $T \gtrsim \Delta_0$, the conductance reaches a plateau, which can be significantly lower than $2e^2/h$, especially if the interaction is extremely strong $K_c = 0.1$ (see Fig.~\ref{fig:conductance_frictionK0102}a). The drop in conductance can be as large as $0.5e^2/h$ even in relatively short wires $L \gtrsim 5\hbar v_F/\Delta_0$. One can numerically estimate the corresponding length as $L\gtrsim 0.5\;\mathrm{\mu m}$ for $\Delta_0\sim 0.5\; \mathrm{meV}$.
 However, even in a more realistic case of a weaker interaction strength $K_c=0.2$, the drop of the conductance at finite temperature is still large to be observed experimentally if the wire is long enough (Fig.~\ref{fig:conductance_frictionK0102}b). 
 In this case the drop of conductance can reach $0.1e^2/h$ for much longer wires $L\gtrsim 50\hbar v_F/\Delta_0$, corresponding to a length $L\gtrsim 5\;\mathrm{\mu m}$ for $\Delta_0\sim 0.5\; \mathrm{meV}$.

\section{Conclusions \label{sec:results}}

In this work, we analyzed electrical transport properties  of a quantum wire, in particular the conductance,  in the presence of strong electron-electron interactions inside the wire. Many-particle backscattering processes caused by electron-electron interactions lead to the formation of a fractional Luttinger liquid state with a partial gap in the spectrum. Using bosonization and LL formalism, we studied how the gap manifests itself in the conductance and how it is affected by the presence of a smoothly varying gate potential $V_g(x)$ determining the connection between wire and leads. We analyzed this problem in two complementary approaches, one where we solve the problem essentially exactly but for a special value of the interaction strength, allowing refermionization, and
a second one, which is based on a semiclassical approach but valid for arbitrary interaction strengths.
As an important result, we found that even if the chemical potential lies inside the partial gap but is not close enough to the resonance value, the fractional conductance cannot be observed. This means that an experimental observation of  fractional behavior requires a rather precise fine-tuning of the system parameters. 

We also predict a mechanism of resistivity caused by the interaction of the sine-Gordon solitons with gapless fluctuation modes. This mechanism leads to a suppression of the conductance at finite temperatures, and also leads to a dependence of the conductance on the length of a long and clean quantum wire and on the strength of the electron-electron interactions in stark contrast to the case of the quantized conductance of conventional Luttinger liquids. Thus, in order to observe this effect experimentally one needs to probe a sufficiently long (with the length of several micrometers) clean wire with strong electron-electron interactions.

\section*{Acknowledgments}
We would like to thank Leonid Glazman and Pascal Simon for helpful discussions. This work was supported by the Swiss National Science Foundation (Switzerland) and by the NCCR QSIT. This project has received funding from the European Union’s Horizon 2020 research and innovation program (ERC Starting Grant, grant agreement No 757725). 

\appendix

\section{RG analysis\label{app:RG}}

\begin{figure}
	\includegraphics[width=\columnwidth]{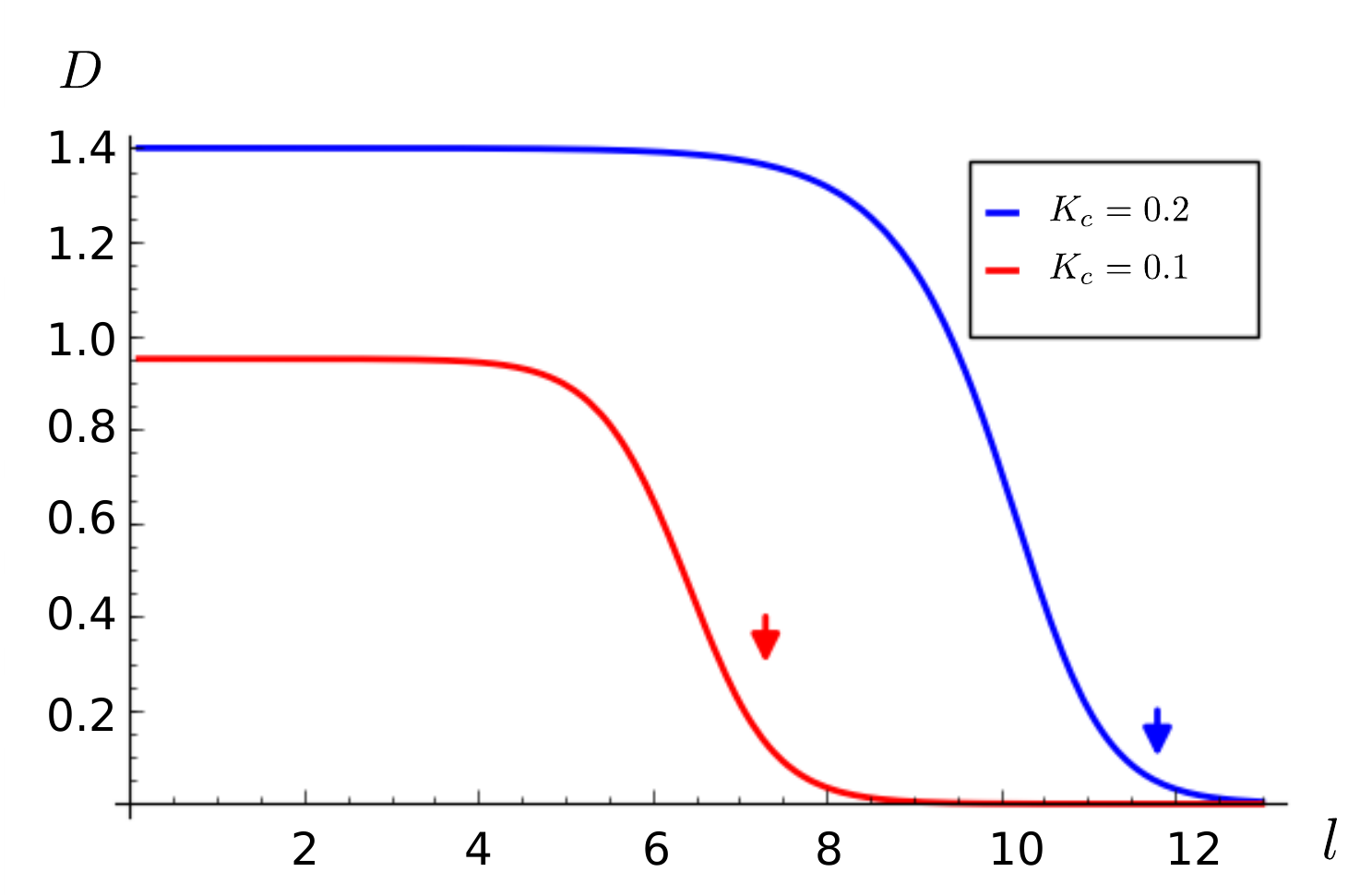}
	\caption{The RG flow of the scaling dimension $D(l)$ of the sine-Gordon term in the action $S_1$ [see Eq.~(\ref{eqn:action-gapped})] obtained by  numerical solution of the RG equations [see ~Eqs.~(\ref{eqn:RG-1})--(\ref{eqn:RG-4})] for two different values of Luttinger liquid parameter $K_c$, which are chosen to be small enough so that the sine-Gordon term $\omega_0^2\cos\left(\sqrt{2}\lambda \phi_1\right)$ in~Eq.~(\ref{appeqn:action-gapped}) generating the gap is relevant.	 As the flow parameter grows, the scaling dimension of the sine-Gordon term becomes small, $D\ll 1$, indicating strong pinning of the field $\phi_1$. Although the initial scaling dimension is lower for a stronger interaction $K_c=0.1$, the scaling dimension at the point where the flow stops (shown with vertical arrows) is smaller for $K_c=0.2$. For initial values we took $z=10^{-3}$ and velocities for a corresponding value of interaction parameter $K_c$ from Table~\ref{table:velocities}.}
	\label{fig:D-flow}

\end{figure}

In this subsection, we study the action $S=S_1 + S_2 + S_{12}$ given by Eqs.~(\ref{eqn:action-gapped})--(\ref{eqn:action-coupling}) in the following form, where all the parameters are expressed via $v_{1,2,12}$ as well as  $K_{1,2}$ and, for the sake of simplicity, their spatial dependence is disregarded:
	\begin{align}
\begin{split}
S_1 &=\; -\frac{1}{2\pi v_1 K_1}\int dxd\tau\; \left\{ \phi_1\left[  -\partial_\tau^2 -  v_1^2\partial_x^2 \right]\phi_1\right.\\ &\left.
+\omega_0^2\cos \left(\sqrt{2}\lambda \phi_1\right)\right\},
\end{split}
\label{appeqn:action-gapped}
\\
S_2 &=\; -\frac{1}{2\pi v_2 K_2}\int dxd\tau\;  \phi_2\left[  - \partial_\tau^2  -   v_{2}^2 \partial_x^2  \right]\phi_2,\\
S_{12} &=\; -\frac{1}{2\pi \sqrt{v_1 K_1 v_2 K_2}}\int dxd\tau\;  \phi_1\left[ 2 v_{12}^2 \partial_x^2 \right]\phi_2.	
\label{appeqn:action-coupling}
\end{align}
 We use  standard  RG techniques~\cite{GiamarchiBook}, and treat the sine-Gordon term $\omega_0^2\cos \left(\sqrt{2}\lambda \phi_1\right)$  as a perturbation. The renormalization of velocities ($v_{1}$, $v_{2}$, $v_{12}$), of Luttinger liquid parameters ($K_{1}$, $K_{2}$), and of the gap $\Delta$ is described by the following RG equations:
\begin{align}
&\frac{dz}{dl} = (2-D)z,\label{eqn:RG-1}\\ 
&\frac{d\left(v_{1}/K_{1}\right)}{dl} = \frac{K_{1} v_{1} v_{12}^2}{u_1^2 u_2^2}v_2^2z^2,\\
&\frac{d\left(\frac{1}{v_{1}K_{1}} \right)}{dl} = -\frac{1 - v_2^2\left(u_1^{-2}+u_2^{-2} \right)}{u_1^2 u_2^2}K_{1} v_{1}v_2^2z^2,
\label{eqn:RG-3}\\
&\frac{d\left(v_{12}^2/\sqrt{v_{1}K_{1}} \right)}{dl} = 0,\\
&\frac{d K_{2}}{dl} =0,\ \  \frac{d v_{2}}{dl} = 0,
\label{eqn:RG-4}
\end{align}
where we have introduced the flow parameter $l = \ln\left[a/a_0\right]$. The dimensionless coupling constant is defined as $z(l) = \Delta(l) a(l)/v_2$ and its scaling dimension $D(l)$ is given by  
\begin{align}
D = \frac{\left(\gamma_c^2 + 1\right)\left(u_1u_2 + v_{2}^2 \right)K_{1}v_{1}}{2(u_1+u_2)u_1u_2}.
\end{align}
Here, $u_{1,2}$ are the velocities of the eigenmodes of the quadratic part of the action, which are given by the positive roots of the characteristic equation,
\begin{align}
u^4 - (v_{1}^2 + v_{2}^2)u^2 + v^2_{1}v^2_{2} - v_{12}^4 = 0.
\end{align}
The trivial integrals of the RG flow equations [see ~Eqs.~(\ref{eqn:RG-1})--(\ref{eqn:RG-4})]  are $K_{2}$, $v_{2}$, $J_1 = v_{12}^2/\sqrt{v_{1}K_{1}}$. However, it can be shown straightforwardly that $J_2 = u_{1}^{-2} + u_{2}^{-2} = \left(v_{1}^2 + v_{2}^2\right)/(v_{1}^2 v_{2}^2 - v_{12}^4)$ is also a first integral of the system of coupled equations,
 whose value can be determined from the initial conditions as
\begin{align}
J_2 = v_c^{-2} + v_\sigma^{-2} = \left(K_c^2 + 1\right)/v_F^2\, .
\label{J2}
\end{align}

\begin{figure}
	\includegraphics[width=\columnwidth]{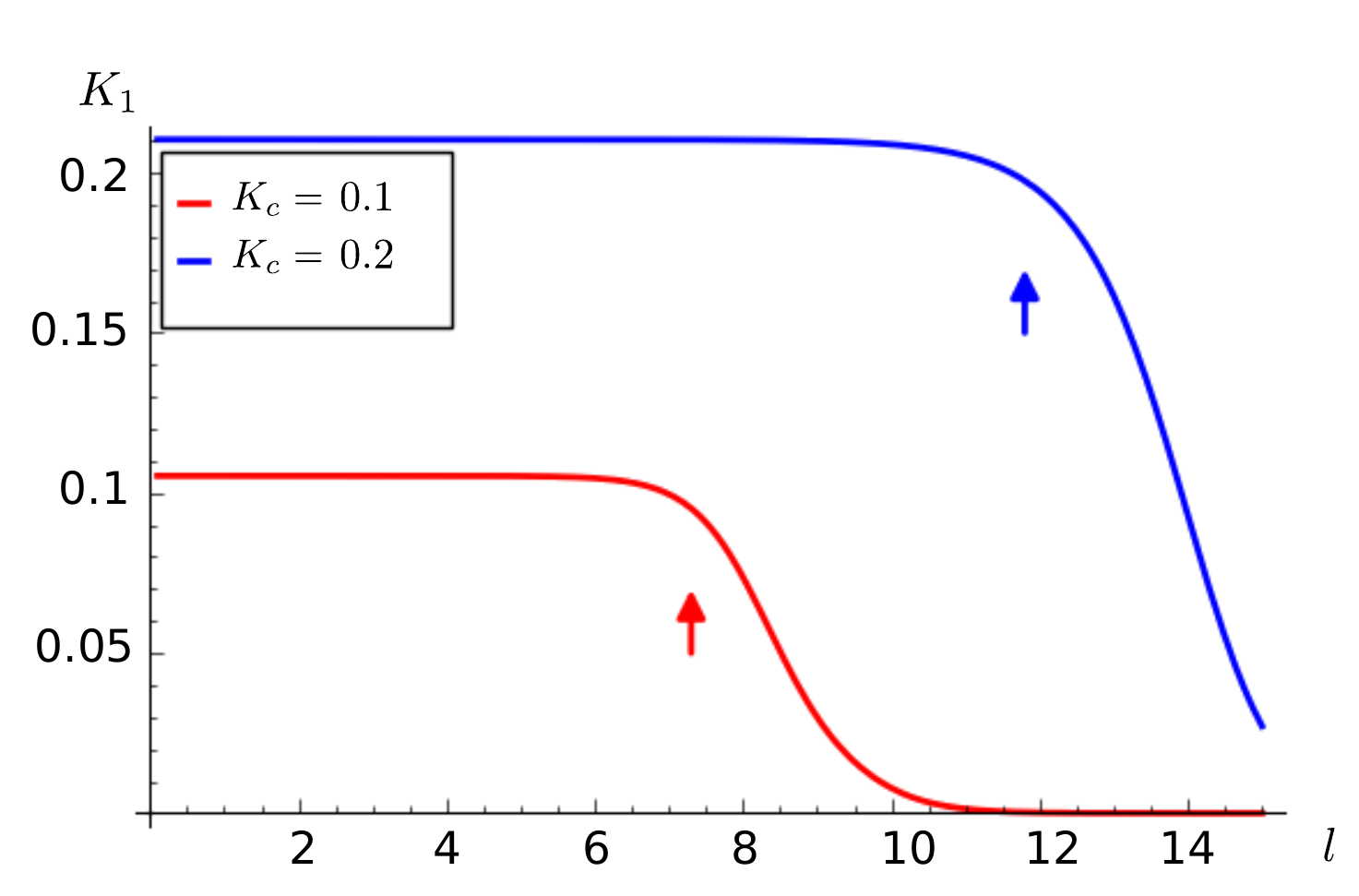}
	\caption{
		The same as Fig.~\ref{fig:D-flow} but for the flow of the interaction parameter $K_1(l)$. Although, the interaction parameter decreases during the flow, it does not change significantly.}
	\label{fig:K1-flow}
\end{figure}

\begin{figure}
	\includegraphics[width=\columnwidth]{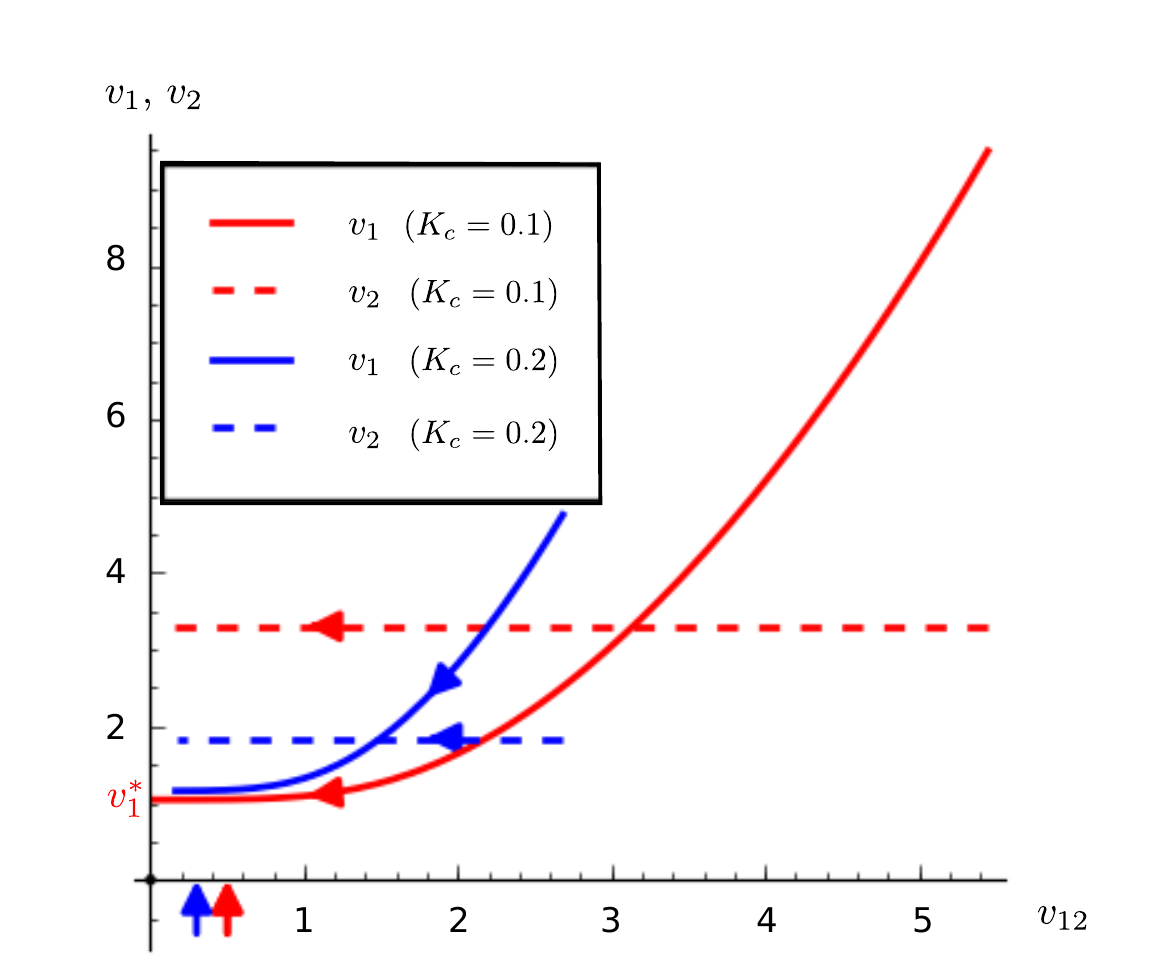}
	\caption{The RG flow of velocities $v_{1}$ (solid lines) and $v_{2}$ (dashed lines) as functions of $v_{12}$. The initial values of Luttinger liquid parameter is chosen as $K_c=0.2$ (blue lines) and $K_c=0.1$ (red lines). All velocities are plotted in units of $v_F$ and vertical arrows show a point at which the flow is stopped.	 First, $v_2$ remains constant during the RG flow in accordance with RG equations [see ~Eq.~(\ref{eqn:RG-4})]. 
Second, $v_1$ flows to a value $v_1^{*}$ close to $v_F$ (at $K_c=0.1$ and $K_c=0.2$ the corresponding values of $v_1^*$ are $v_1^* \approx 1.04 v_F$ and $v_1^* \approx 1.2 v_F$, respectively). Whereas, an amplitude  of the cross-term $v_{12}$ in the action $S_{12}$ [see Eq.~(\ref{appeqn:action-coupling})] flows to a much smaller value, $v_{12} < v_1$, which means that the cross term given by Eq.~(\ref{appeqn:action-coupling}) can be treated as a perturbation.
}
	\label{fig:velocities-flow}
\end{figure}

For strong electron-electron interactions, $K_c < 3/\gamma_c^2$, the sine-Gordon term is relevant [see discussion above~Eq.~(\ref{eqn:1st-RG})]. Its scaling dimension $D (l)$ and the effective LL parameter for the first mode $K_{1}$ decrease with increasing  RG flow parameter  $l$, see~Figs.~\ref{fig:D-flow} and \ref{fig:K1-flow}, respectively. We note that, in case of a sufficiently long wire and in the limit of zero temperature, the RG flow should be stopped at  $z(l) \sim 1$, meaning that the flowing cut-off parameter $a(l)$ reaches the correlation length $v_F/\Delta$. 
Using the fact that $J_1$ and $J_2$ are first integrals, we can come to the conclusion that $v_{12}$ should flow to zero (to be accurate, the flow is stopped at $z(l)\sim 1$, and $v_{12}$ remains finite but small, $v_{12}\ll v_F$), while $v_{1}$ flows to a finite value (see~Fig.~\ref{fig:velocities-flow}), which can be determined from Eq. (\ref{J2}) as
\begin{align}
\left(v_{1}^*\right)^{-2} = \frac{K_c^2 + 1}{v_F^2} - v_{2}^{-2}.
\end{align}
As a consequence, at large values of $l$, $v_{1}(l), v_{2}(l) \gg v_{12}(l)$. This means that the cross term $S_{12}$, defined in Eq.~(\ref{eqn:action-coupling}), can be treated as a small perturbation.

\section{Calculating transport properties in refermionization approach\label{app:refermionization}}

\subsection{Contribution to the conductance from the states inside the gap}

In this subsection, we solve the set of equations of motion [see Eqs.~(\ref{eqn:refermion-motion-0})--(\ref{eqn:refermion-motion-1})] for the refermionized fields $R_1$ and $L_1$. For the states with energies inside the gap $|\varepsilon| < \Delta$, the general form of the solution is defined in Eq.~(\ref{eqn:solution-below-gap}). The expectation value of the tunneling current $j_1$ carried by gapped refermions can be expressed via the correlators of operators $A$, $B$ [see Eq.~(\ref{eqn:solution-below-gap})] as
\begin{align}
\langle j_1 \rangle = s \left(\langle A^\dag B \rangle \frac{\kappa}{\kappa-iq} + c.c. \right)e^{-\kappa L},
\label{eqn:j1-via-AB}
\end{align}
where $q= \varepsilon/s$ [cf. Eq.~(\ref{eqn:refermion-tunneling-current})]. We also express densities of the gapped refermions at the left ($\rho_{1,L}$) and the right ($\rho_{1,R}$) leads as
\begin{align}
\langle\rho_{1,L}\rangle &=\; \langle A^\dag A \rangle + e^{-2\kappa L} \langle B^\dag B \rangle +
e^{-\kappa L}\varepsilon \left( \frac{\langle A^\dag B \rangle}{\varepsilon - i\kappa s} + c.c. \right),\\
\langle\rho_{1,R}\rangle &=\; \langle A^\dag A \rangle e^{-2\kappa L} + \langle B^\dag B \rangle +
e^{-\kappa L}\varepsilon \left( \frac{\langle A^\dag B \rangle}{\varepsilon - i\kappa s} + c.c. \right).
\end{align}
In addition, from the boundary conditions defined in Eq.~(\ref{eqn:refermion-boundary}), we obtain
\begin{align}
&\left\langle A^\dag A + B^\dag B \right\rangle \frac{\varepsilon}{\Delta}e^{-\kappa L} \nonumber\\
& \hspace{30pt}=\frac{1+e^{-2\kappa L}}{2}\left( \langle A^\dag B \rangle \frac{\varepsilon+i\kappa s}{\Delta} + c.c. \right),\\
&\left\langle  B^\dag B - A^\dag A \right\rangle \frac{\kappa s}{\Delta}e^{-\kappa L} \nonumber\\
& \hspace{30pt}= \frac{1-e^{-2\kappa L}}{2}\left( i\langle A^\dag B \rangle \frac{\varepsilon+i\kappa s}{\Delta}+  c.c. \right).
\label{eqn:bc-via-AB}
\end{align}
Finally, solving Eqs.~(\ref{eqn:j1-via-AB})--(\ref{eqn:bc-via-AB}) together, we arrive at
\begin{align}
&\langle A^\dag A \rangle = \rho_{1,L} + \frac{1}{2}(\rho_{1,L} - \rho_{1,R})e^{-\kappa L}\frac{\varepsilon^2}{\Delta^2} +  O\left( e^{-2\kappa L} \right), \nonumber\\
&\langle B^\dag B \rangle = \rho_{1,R} + \frac{1}{2}(\rho_{1,R} - \rho_{1,L})e^{-\kappa L}\frac{\varepsilon^2}{\Delta^2} +  O\left( e^{-2\kappa L} \right),\nonumber\\
& \langle j_1 \rangle = 2s\frac{\kappa^2 s^2}{\Delta^2}\left\langle \rho_{1,R} - \rho_{1,L} \right\rangle e^{-2\kappa L} +  O\left( e^{-3\kappa L} \right),
\label{eqn:j1-app}
\end{align}
where we have kept only the leading terms for simplicity.

Substituting~Eq.~(\ref{eqn:j1-app}) and Eqs.~(\ref{eqn:refermion-relation1})--(\ref{eqn:refermion-relation4}) into the boundary conditions~Eqs.~(\ref{eq:Egger-Grabert-charge-L})--(\ref{eq:Egger-Grabert-spin-R}), we obtain a system of four independent linear equations for the variables $\rho_{1,L}$, $\rho_{1,R}$, $\rho_{2}$, $j_2$, which can be solved straightforwardly. In  leading order we obtain
\begin{align}
\langle\rho_{1,R} - \rho_{1,L}\rangle = \frac{\sqrt{2} \lambda \gamma_c \Xi}{v_F\left(1 + \gamma_c v_{12}^2/v_2^2 \right)} \frac{n_R-n_L}{\pi\left(\gamma_c^2 +1\right)} + O(e^{-2\kappa L}).
\label{eqn:rho-difference}
\end{align}
Combining Eq.~(\ref{eqn:rho-difference}) with Eq.~(\ref{eqn:j1-app}), we obtain a tunneling contribution to the current $j_1 \propto e^{-2\kappa L}$ from the gapped modes. Substituting it to Eq.~(\ref{eqn:refermion-relation2}), we arrive at Eq.~(\ref{eqn:delta-G}) given in the main text.

\subsection{Contribution to the conductance from the states above the gap}

Refermions with energies above the gap $\varepsilon>\Delta$ can be treated in the same way as discussed in the preceding subsection. The general solution of Eqs.~(\ref{eqn:refermion-motion-0})--(\ref{eqn:refermion-motion-2}) can be represented as
\begin{align}
\begin{pmatrix}
R_1\\
L_1
\end{pmatrix}
= 
A
\begin{pmatrix}
u_k\\v_k
\end{pmatrix} e^{ikx}
+
B
\begin{pmatrix}
v_k\\u_k
\end{pmatrix} e^{-ik(x-L)},
\end{align}
where $k = \sqrt{\varepsilon^2 - \Delta^2}/s$, $u_k = \sqrt{1/2 + k/2q }$, 
$v_k = \sqrt{1/2 - k/2q }$. Here, $A$ and $B$ are again fermionic operators representing the right- and left-moving refermions.
The expectation values of current  $j_1$ as well as of the densities $\rho_{1,L}$ and $\rho_{1,R}$ of the gapped refermions   can be expressed as
\begin{align}
&\langle j_1 \rangle/s = \frac{\sqrt{\varepsilon^2- \Delta^2}}{\varepsilon^2}
\left\langle A^\dag A - B^\dag B \right\rangle, \\
&\langle \rho_{1,L} \rangle = \langle A^\dag A \rangle + \langle B^\dag B \rangle
+\frac{\Delta}{\varepsilon}\left( \langle A^\dag B \rangle e^{ikL} + c.c. \right),\nonumber \\
&\langle \rho_{1,R} \rangle = \langle A^\dag A \rangle + \langle B^\dag B \rangle
+\frac{\Delta}{\varepsilon}\left( \langle A^\dag B \rangle e^{-ikL} + c.c. \right).\nonumber 
\end{align}
The boundary conditions given by Eq.~(\ref{eqn:refermion-boundary}) can be rewritten as
\begin{align}
\frac{\Delta}{\varepsilon}\langle A^\dag A + B^\dag B \rangle\cos kL =
\langle A^\dag B \rangle \left(u_k^2 +v_k^2e^{-2ikL} \right) + c.c.,\\
\frac{\Delta}{\varepsilon}\langle A^\dag A + B^\dag B \rangle\sin kL =
i\langle A^\dag B \rangle \left(u_k^2 -v_k^2e^{-2ikL} \right) + c.c.
\end{align}
Again, similarly to the previous subsection, $j_1$ can be expressed as a linear function of $\rho_{1,R}$ and $\rho_{1,L}$. Using the boundary conditions defined in Eqs.~(\ref{eq:Egger-Grabert-charge-L})--(\ref{eq:Egger-Grabert-spin-R}), we obtain a system of linear equations for $\rho_2$, $j_2$, $\rho_{1,L}$, $\rho_{1,R}$, which we solve numerically. Expressing $j_c$ with the help of Eq.~(\ref{eqn:refermion-relation2}), we obtain numerically  the effective transmission shown in  Fig.~\ref{fig:transmission}. The numerical calculations to obtain Figs.~\ref{fig:conductance-T0},~\ref{fig:conductance-T}--\ref{fig:conductance-profiles}, as well as Figs.~\ref{fig:G-vs-mu-adiabatic}--\ref{fig:G-vs-mu-abrupt}, are performed along the same lines.

\section{Effective action for one kink \label{app:eff-action}}
In this subsection, we follow~Refs.~\onlinecite{BraunPRB1996, Aseev2017} and derive an effective action for the kink-particle up  to order $O\left( (\partial_\tau\xi/v_F) ^2\right)$. The action $S_E = S_1 + S_2 + S_{12}$ [ see~Eqs.~(\ref{eqn:action-gapped})--(\ref{eqn:action-coupling})] can be expanded around the one-kink solution,
$\check{\varphi}(x,\tau) = \check{\phi}^{k}(x-\xi) + \delta\check{\varphi}(x-\xi(\tau),\tau)$, where $\delta\check{\varphi}(x,\tau)$ describes fluctuations around the  static classical  path, 
\begin{align}
&S= S_\xi + S_\varphi,\\ 
&S_\xi = \frac{1}{2\pi v_F}\int d\tau\; \left[\int dx\; (\partial_x\check{\phi}^{K},{\partial_x\check{\phi}^{K}})\right](\partial_\tau
\xi)^2,\label{eqn:Sxi}\\
&S_\varphi =  \frac{1}{2\pi v_F}\int dx d\tau\; \left(\delta\check{{\varphi}},({\mathscr{H}}_0 + {\mathscr{H}_1})  \delta\check{\varphi}\right) + \left(\check{\mathscr{J}}, \delta\check{{\varphi}} \right), \label{WE}
\end{align}
where we use the following notation for the scalar products: $\left(\mathscr{J}, \delta\check{{\varphi}} \right)={\mathscr{J}}_1  \delta{{\varphi}_1} + {\mathscr{J}}_2  \delta{{\varphi}_2}$ and
$
\left(\partial_x \check{\phi}^K, \partial_x \check{\phi}^K\right) = \left(\partial_x \check{\phi}^K\right)^T\partial_x \check{\phi}^K.
$ 
The operators $\mathscr{H}_0$, $\mathscr{H}_1$, and the spinor $\mathscr{\check{J}}$ are defined as 
\begin{align}
&{\mathscr{H}_0} = \begin{pmatrix}
\partial_\tau^2&0\\
0&\partial_\tau^2
\end{pmatrix} +\partial_x\begin{pmatrix}
v_{1}^2&v_{12}^2\\
v_{12}^2&v_{2}^2
\end{pmatrix}\partial_x - \begin{pmatrix}
V^2(x)&0\\
0&0
\end{pmatrix},\\
& {\mathscr{H}}_1 =   \left[2(\partial_\tau \xi) \partial_x\partial_\tau - (\partial_\tau \xi)^2\partial_x^2\right]\begin{pmatrix}
1&0\\
0&1
\end{pmatrix} ,\\
&\check{\mathscr{J}} = -2\left(\partial_\tau \xi\right)^2 \partial_x^2\check{\phi}^K,
\end{align}
and the potential $V^2(x)$ is given by
\begin{align}
V^2(x) = \omega_0^2\left(1 - 2\mathrm{sech}^2\left(x/\delta_0\right) \right).
\end{align}  
The effective action for the collective coordinate $\xi$ can be represented as
\begin{align}
S_{eff}[\xi] = S_\xi -\ln\left\{\int^\prime \mathcal{D}\delta\varphi\; \det \left(\frac{\delta Q}{\delta \xi} \right)e^{-S_\varphi} \right\}.
\label{eqn:eff-action}
\end{align}  
The prime denotes that the  integration is performed over fluctuations orthogonal to the zero-mode $\partial_x\check{\phi}^k$.
In order to integrate out fluctuations, we first eliminate a linear term in Eq. (\ref{WE}) by shifting $\check{\varphi}$ by $\check{\rho} \equiv (1/2)\mathscr{H}^{-1}\mathscr{J}$, $\check{\varphi} \to \check{\varphi} - \check{\rho}$.  Similar to~Refs.~\onlinecite{BraunPRB1996, Aseev2017}, the Faddeev-Popov (Jacobian) determinant $\det (\delta Q/\delta \xi)$ leads to an extra term in the action proportional to $(\partial_\tau \xi)^2$, which results in a mass renormalization. Its exact value is not of interest here since we can estimate the value of the renormalized mass $M=\Delta/s^2$ by using~Eq.~(\ref{eqn:gap}).

We now turn to the integration over $\delta\check{\varphi}$ in~Eq.~(\ref{eqn:eff-action}),
\begin{align}
\int^\prime \mathcal{D}\delta\varphi\; e^{-S_\varphi} = \frac{1}{\sqrt{\det'\left(\mathscr{H}_0+\mathscr{H}_1 \right)}}.
\end{align}
The prime on the determinant denotes omission of the zero mode $\partial_\xi \check{\phi}^K(x,\xi)$. Using the identity $\ln \det = \mathrm{tr} \ln$, we expand
\begin{multline}
\frac{1}{\sqrt{\det'\left(\mathscr{H}_0+\mathscr{H}_1 \right)}} = \exp\left\{ -\frac{\mathrm{tr}' \ln\left(\mathscr{H}_0\left[ 1+\mathscr{H}_0^{-1}\mathscr{H}_1\right] \right)}{2} \right\}\\
\approx \frac{1}{\sqrt{\det' \mathscr{H}_0}} \exp\left\{ -\frac{\mathrm{tr}' \left[ \mathscr{H}_0^{-1}\mathscr{H}_1 -\frac{1}{2}\left(\mathscr{H}_0^{-1}\mathscr{H}_1\right)^2\right] }{2} \right\}.
\end{multline}
Since $\mathscr{H}_1 = O\left( \partial_\tau \xi/v_F\right)$, this expression represents an expansion in increasing powers of $\partial_\tau \xi/v_F$. When fluctuations $\delta \check{\varphi}$ are integrated out, the action becomes quadratic in the collective coordinate $\xi$.
Similar to~Refs.~\onlinecite{BraunPRB1996, Aseev2017}, the first order term $\mathscr{H}_0^{-1}\mathscr{H}_1$  leads to terms proportional to $\partial_\tau \xi^2$, which again result in the mass renormalization. As we will show, the second order term $\left(\mathscr{H}_0^{-1}\mathscr{H}_1\right)^2$ generates an additional term, which is non-local in time and which results in  internal friction. 

The operator ${\mathscr{H}_0}$ describes free mesons. The spectrum of mesons can be found by solving the Schr\"odinger equation for the eigenfunctions far away from the kink center, where the scattering barrier $V^2(x)$ created by the kink vanishes,
\begin{align}
\begin{pmatrix}
v_{1}^2q^2 + \omega_0^2 - \omega_q^2 &v_{12}^2q^2\\
v_{12}^2q^2&v_{2}^2q^2-\omega_q^2
\end{pmatrix}\delta\check{\varphi}_q = 0  .
\end{align}
The fluctuation spectrum consists of two branches, which we will denote by index $\nu=\pm$. One of them ($\nu=+$) has a gap $\omega_0$ and the other ($\nu=-$) is gapless,
\begin{align}
\left(\omega_q^{\nu=\pm}\right)^2 = \frac{\omega_0^2+2 q^2 v_{1}^2\pm\sqrt{\left[\omega_0^2 + \left(v_{1}^2-v_{2}^2 \right)q^2 \right]^2 + 4v_{12}^4q^4 }}{2}.
\end{align}
At low momentum $q\ll \omega_0/ v_{1}$, the gapped branch $\omega^+_q$ and the gapless branch $\omega^-_q$ are described  by
\begin{align}
\omega_q^+ \approx \sqrt{\omega_0^2 + v_{1}^2 q^2},\quad \omega_q^- \approx v_{2}|q|.
\end{align} 
The eigenfunctions of $\mathscr{H}_0$ factorize into a space and time part $\check{\Phi}_{\nu,q,\bar{\omega}}(x,\tau) = \Phi_{\nu,q}(x)e^{i\bar{\omega}\tau}/\sqrt{\beta}$, where $\beta=1/T$.
 Using these notations, we find, up to the second order in the small parameter $(\partial_\tau \xi/v_F)^2$,
\begin{multline}
\frac{1}{4}\mathrm{tr'}\left(\mathscr{H}_0^{-1}\mathscr{H}\right)^2 \\= 
\sum\limits_{\nu=\pm,q,q',\bar{\omega},\bar{\omega}'}
\frac{\left|\int dxd\tau\; \left(\Phi_{\nu,q',\bar{\omega}'},(\partial_\tau \xi)\partial_\tau{\partial_x} \Phi_{\nu,q,\bar{\omega}}  \right) \right|^2} 
{\left(\bar{\omega}^2+\left(\omega_q^{\nu}\right)^2\right)\left(\bar{\omega}^2+\left(\omega_{q'}^{\nu}\right)^2\right)}.
\label{eqn:trace-2nd-order}
\end{multline}
In leading order in wire length $L$, we get
\begin{multline}
\int dxd\tau\;\left(\Phi_{\nu,q',\bar{\omega}'},(\partial_\tau \xi)\partial_\tau\partial_x \Phi_{\nu,q,\bar{\omega}}  \right)
=\\ -\frac{q\bar{\omega}'}{\beta}\delta_{qq'}\int d\tau
e^{i(\bar{\omega}'-\bar{\omega})\tau}\partial_\tau \xi(\tau).
\end{multline}
Thus, Eq.~(\ref{eqn:trace-2nd-order}) can be rewritten in the form 
\begin{align}
\frac{1}{4}\mathrm{tr'}\left(\mathscr{H}_0^{-1}\mathscr{H}\right)^2 = 
T\sum\limits_{\nu=\pm,\bar{\omega}} \bar{\omega}^2 \xi_{\bar{\omega}} \xi_{-\bar{\omega}} \Gamma^\nu(\bar{\omega}),
\end{align}
where $\xi_{\bar{\omega}} = \int_0^\beta d\tau e^{i\bar{\omega}\tau} \xi(\tau)$ is the Fourier transform of the collective coordinate. The damping kernel $\Gamma^\nu(\bar{\omega})$ is defined as
\begin{multline}
\Gamma^\nu(\bar{\omega}) = T\sum\limits_{\bar{\omega}', q}\frac{q^2\left(\bar{\omega}+\bar{\omega}'\right)\bar{\omega}'}{\left[\left(\bar{\omega}+\bar{\omega}'\right)^2+\left(\omega^{\nu}_q\right)^2\right]\left[\left(\bar{\omega}^{\prime}\right)^2+\left(\omega^{\nu}_{q}\right)^2\right]}.
\end{multline}
Next, performing the summation over bosonic Matsubara frequencies $\bar{\omega}'$, we obtain
\begin{align}
\Gamma^\nu(\bar{\omega}) = \sum\limits_q \frac{4q^2\omega^\nu_q \coth\left(\beta \omega^\nu_q/2\right)}{4\left(\omega_q^{\nu}\right)^2+\bar{\omega}^2}.
\end{align}
To render the results stay finite in the thermodynamic limit, we have to subtract the vacuum fluctuations~\cite{Sakita1985}. This renormalization simply amounts to the replacement~(see~Ref.~\onlinecite{BraunPRB1996} for more details) 
\begin{align}
\sum_{\nu,q}\to \sum_{\nu}\int dq \left[\rho^\nu(q) - \frac{L}{2\pi} \right],\\
\rho^+_q - \frac{L}{2\pi} \approx  \frac{2 v_{1}^2 }{\left(\omega_q^+\right)^2 \delta_0 }, \\
\rho^-_q - \frac{L}{2\pi} \approx  \frac{2 v_{12}^2  }{\left(\omega_q^-\right)^2 \delta_0 }\, ,
\end{align}
where $\rho^\nu(q)$ is the density of states for the gapped ($\nu=+$) and gapless ($\nu=-$) modes, respectively. Note that in the limit when the coupling $v_{12}^2$ between gapless mode and kink vanishes, the density of states for the gapless mode $\rho^-_q$ is not affected by scattering at the kink,  $\rho^-_q = L/(2\pi)$.  The damping kernels $\Gamma^\nu$ are then given by
\begin{align}
\Gamma^+(\bar{\omega}) = \int \frac{dq}{2\pi} \frac{2\delta_0^{-1} v_{1}^2 }{\left(\omega_q^{+}\right)^2}\frac{4q^2\omega^+_q \coth\left(\beta \omega^+_q/2\right)}{4\left(\omega_q^{+}\right)^2+\bar{\omega}^2},\\
\Gamma^-(\bar{\omega}) = \int \frac{dq}{2\pi} \frac{2\delta_0^{-1} v_{12}^2 }{\left(\omega_q^{-}\right)^2}\frac{4q^2\omega^-_q \coth\left(\beta \omega^-_q/2\right)}{4\left(\omega_q^{-}\right)^2+\bar{\omega}^2}.
\end{align}

First, for the gapped mode, the integration for $\Gamma^+$ does not diverge in the infrared limit, and $\Gamma^+$ is of order $O(\bar{\omega}^0)$. Therefore, the gapped modes contribute only to the mass (gap) renormalization.

Second, in order to estimate $\Gamma^-$, we linearize the spectrum of gapless fluctuation modes $\omega_q^- \approx v_{2}q$, since the main contribution to the integral is in the limit of small momenta $q$. The integration yields
\begin{align}
\Gamma^-(\bar{\omega}) = 2 \frac{\delta_0^{-1}v_{12}^2}{v_{2}^3}\frac{T}{|\bar{\omega}|} + O(\bar{\omega}^0).
\end{align}
The resulting effective action for the collective coordinate is now given by
\begin{align}
S_{eff}[\xi] = T\sum\limits_{\bar{\omega}} \left[\frac{M}{2}\bar{\omega}^2 + 2 \frac{\delta_0^{-1}v_{12}^2}{v_{2}^3} T |\omega|  \right]\xi_{-\bar{\omega}}\xi_{\bar{\omega}}.
\end{align}
The first term, stemming from Eq.~(\ref{eqn:Sxi}),
  describes the kinetic energy of a particle with a mass $M = \Delta/s^2$, which can be determined by using Eq.~(\ref{eqn:1st-RG}). The second term describes Ohmic friction experienced by a kink coupled to a bath of gapless mesons. This friction term is temperature-dependent and vanishes at zero temperature.


\begin{thebibliography}{99}
\bibitem[]{footnote1}{
	Here for convenience we omit the units for the electron charge and spin, $e$ and  $\hbar/2$, respectively, in  the definitions of $\rho_{c,\sigma}$ and $j_{c,\sigma}$. The complete definitions read, 
	$\rho_c = -\sqrt{2}e\partial_x \phi_c / \pi$, $\rho_\sigma = \hbar \partial_t \theta_\sigma/(\sqrt{2}\pi v_F)$ $j_c = \sqrt{2}e\partial_t \phi_c/\pi$, $j_\sigma = -\hbar v_F\partial_x \theta_\sigma/(\sqrt{2}\pi)$.
}
\end{thebibliography}
\end{document}